\documentstyle{article}
\begin{document}
\begin{titlepage}

YCTP-P38-91  October 24, 1991

\bigskip

\centerline{POLYMERS AND PERCOLATION IN TWO DIMENSIONS}
\centerline{AND TWISTED N=2 SUPERSYMMETRY}
\vspace{0.5cm}
\centerline{H.Saleur\footnote{On leave from SPHT Cen Saclay 9119 Gif Sur Yvette
Cedex France}\footnote{Work supported in part by DOE grant DE-AC0276ER03075
 and by
a Packard Fellowship for Science and Engineering}}
\centerline{Physics Department}
\centerline{Yale University}
\centerline{New Haven CT 06519}
\centerline{USA}
\vspace{1cm}
\centerline{Abstract}
\vspace{0.5cm}
We show  how a large class of  geometrical critical systems including
dilute
polymers, polymers at the theta point,
 percolation and to some extent brownian motion,
are described by a twisted N=2 supersymmetric theory with $k=1$
 (it is broken in the dense
polymer phase that is described simply by  a $\eta,\xi$ system).
 This allows us to give for the first time a
consistent conformal field theory description of these problems. The fields
that were described so far by formally allowing half integer labels in
the Kac table are built and their four point functions studied. Geometrical
operators are organized in a few representations of the twisted $N=2$ algebra.
A
 noticeable feature is that in addition to Neveu Schwartz
 and Ramond, a sector
with quarter twists has sometimes to be introduced. The algebra of geometrical
operators  is determined.  Fermions boundary conditions are
 geometrically interpreted, and the partition
functions that were  so far defined formally as generating functions for  the
critical exponents are naturally understood, sector by sector.

 Twisted $N=2$ provides moreover a very unified description of all
these geometrical models, explaining for instance why the exponents of polymers
and percolation coincide. It must be stressed that the physical states in these
geometrical problems are {\bf not} the physical states for string theory, which
are usually extracted by the BRS cohomology. In polymers for instance,
$Q_{BRS}$
    is
precisely the operator that creates polymers out of the vacuum, such that the
topological sector is the sector without any polymers.

It seems that twisted $N=2$ is  the correct continuum limit (in two dimensions)
 for models with Parisi Sourlas
supersymmetry. Some possible explanation of this fact is advanced.

As an example of application of $N=2$ supersymmetry we discuss in the first
appendix the still unsolved problem of backbone of percolation. We conjecture
in particular the value $D=25/16$ for the fractal dimension of the backbone, in
good agreement with numerical computations.

Finally some of these ideas are extended to the off critical case in the
second appendix where it is shown how to give a meaning to the $n\rightarrow
0$ limit of the $O(n)$ model  $S$ matrix recently introduced
 by Zamolodchikov
 by introducing fermions.
\end{titlepage}

\section{Introduction}

Since the beginnings of conformal invariance in two dimensions,
geometrical systems have been the subject of active studies. Besides their
interest for computer simulations and experiments, they provide a very
attractive playground for the theory, and have led to predictions of infinite
hierarchies of critical exponents, universal ratios, winding angle
distributions...Despite the remarquable success of these predictions, the
corresponding conformal theories were not so far fully understood. A good
 example
is the dilute polymers where half of the critical exponents, including the
exponent that counts configurations of an open chain, were obtained by using
formally Kac formula with half integer indices for the theory with central
charge $c=0$. If one introduces the $L$ legs
operators $\Phi_{L}$ (figure 1) then several distinct arguments suggest they
are Virasoro primaries with dimension
dimension \cite{SA86}
\begin{equation}
h_{L}=\frac{9L^{2}-4}{96}
\end{equation}
These coincide formally with the Kac dimensions $h_{L/2+2,3}$, but the labels
are integer for $L$ even only. For $L$ odd, the labels are half integer, and
 four point
functions cannot be built in the Dotsenko Fateev Coulomb gas, neither operator
products calculated\footnote{This is discussed in details in \cite{D86}}.
Anothe
   r difficulty is that dilute polymers,
corresponding to  the $n\rightarrow 0$ limit of the $O(n)$ model,
 have a (naive)
partition function equal to one. Some generating functions of critical
dimensions were introduced on formal grounds in \cite{DFSZ87},
 but their meaning remained
so far obscure.

 Similar difficulties are met in the study of  polymers at the theta point, the
percolation problem, or brownian motion.

Still  another kind of problem occurs with dense polymers.
Although it is known they correspond to the $n\rightarrow 0$ limit of the
low temperature phase of the $O(n)$
model, the nature of the broken symmetry and the $"O(-1)"$
Goldstone modes remain mysterious \cite{SA87}.

It is the purpose of this paper to solve the above difficulties. Our result is
that twisted $N=2$ supersymmetry \cite{EY90}
describes correctly all the above models but the dense
polymers. For the latter this symmetry is precisely broken, and they
are described by a $\eta,\xi$ system.  We  show how to determine the
 four point functions
and operator algebra including the fields with half integer labels. We define
various partition functions classified by the fermions boundary conditions. The
trivial $Z=1$ corresponds to doubly periodic sector, while the objects
introduced in \cite{DFSZ87,DS87}
are obtained by summing over Ramond and Neveu Schwartz, and in
some cases over a $Z_{4}$ sector also. The existence of this sector is a
new practical application of the spectral flow \cite{LVW89}.

Although describing geometrical systems by twisted $N=2$ is (so far) merely a
guess justified by results, the idea has its roots in
the work  of Parisi and Sourlas
\cite{PS80}. These authors showed how the $n\rightarrow 0$ limit of the $O(n)$
model could be replaced by  considering  a field theory with $2{\cal N}$ real
bosons $\phi^{i},i=1,\ldots 2{\cal N}$ and $2{\cal N}$ real fermions
$\psi^{i},\overline{\psi}^{i},i=1,\ldots {\cal N}$ such that the Lagrangian is
invariant under the supersymmetry transformations
\begin{equation}
\phi^{i}\rightarrow a_{ij}\phi^{j}+c_{ij}\psi^{j},\psi^{i}\rightarrow
d_{ij}\phi^{j}+b_{ij}\psi^{j}
\end{equation}
that leave ${\bf\phi}^{2}+\overline{\psi}{\psi}$ invariant.  These actually
correspond to a global $Osp(2{\cal N},2{\cal N})$ symmetry. Let us now
 restrict to two
dimensions. In the case of an ordinary compact group $G$, one would expect the
continuum limit of a theory whose lagrangian is
invariant under $G$ to be some Wess Zumino model on
the  group $G$ with local, left right, symmetry \cite{A89}.
 This cannot be expected for
supergroups. One reason is for instance that the polymers can be described by a
$Osp(2{\cal N},2{\cal N})$ symmetric Lagrangian for {\bf any} ${\cal N}$,
 while the WZW models built
on these groups have some features, like the spectrum of dimensions, that
depend on ${\cal N}$  (some other features like the central charge are
 independent of
${\cal N}$, $c=0$ here). A reason for the failure of the usual argument
\cite{A89}
 may be
the non compactness. Also it seems that the introduction of
 fermions and bosons to
simulate a $n=0$ limit leads to somewhat artificial conserved currents that may
not pertain to the correct continuum limit theory. In our case, comparison of
the $Osp(2{\cal N},2{\cal N})$ WZW model spectra with known polymers dimensions
shows that none of these is correct. Although there is no corresponding
lattice model or
field theory a la Parisi Sourlas, there are arguments (see section 4) which
suggest that only one boson and one fermion can propagate along polymer loops.
These degrees of freedom correspond to the fundamental representation of
$Gl(1,1)$ or $Sl(1,1)$. Also these two algebras are the smallest non trivial
subalgebras of the $Osp(2{\cal N},2{\cal N})$ sequence. However one can check
that polymers are not described by a $Gl(1,1)$ WZW model(there is no $Sl(1,1)$
WZW model as discussed in \cite{RS91}).

We have in fact to look for a less naive argument, although the $Gl(1,1)$ WZW
model is "almost" the correct answer. Instead of considering $Gl(1,1)$ we could
start from $Osp(2,2)$ and perform some reduction to keep only the expected
physical currents. Recall the defining relations of $osp(2,2)$ algebra
\[
\left[S_{+},S_{-}\right]=2H
\]
\[
\left[H,S_{\pm}\right]=\pm S_{\pm}
\]
\[
\left[H\pm\frac{I}{2},V_{\pm}\right]=0,\left[H\mp\frac{I}{2}
,V_{\pm}\right]=\pm V_{\pm}
\]
\[
\left[H\mp\frac{I}{2},\overline{V}_{\pm}\right]=0,\left[H\pm\frac{I}{2},
\overline{V}_{\pm}\right]=\pm\overline{V}_{\pm}
\]
\[
\left\{V,V\right\}=\left\{\overline{V},\overline{V}\right\}=0
\]
\[
\left\{V_{+},\overline{V}_{-}\right\}=-\frac{1}{2}(H+\frac{I}{2})
\]
\[
\left\{\overline{V}_{+},V_{-}\right\}=-\frac{1}{2}(H-\frac{I}{2})
\]
\begin{equation}
\left\{V_{\pm},\overline{V}_{\pm}\right\}=\pm\frac{1}{2}J_{\pm}
\end{equation}
For the WZW model on $Osp(2,2)$ all these operators give rise to currents,
whose zero modes satisfy these commutation relations. The stress tensor is
obtained in the usual Sugawara form, and the central charge vanishes.
To keep only the subalgebra $gl(1,1)$ made out of say
$V_{+},\overline{V}_{-},H\pm I/2$\footnote{In the following $gl(1,1)$ is
defined
by a slight change of notation
$\{V_{+},V_{-}\}=E,[F,V_{+}]=F,[F,V_{-}]=-V_{-}$} we must in particular
twist the stress energy tensor such that $c$ remains equal to zero and these
currents keep weight one. The only possible choice is $T\rightarrow
T+\partial(T+H/2)$. Such a modification can be made in two steps, first
twisting $T\rightarrow T+\partial T$, reducing to $N=2$ \cite{BO89},
 and twisting again $T\rightarrow T+\partial(H/2)$ to get twisted $N=2$. For
some unknown reason this procedure is the right one (maybe it is the only
way of getting in the end a theory with spectrum unbounded from below).

The paper is organized as follows. We first discuss dense polymers. It turns
out that a lattice formulation, by means of generalizations
of the Kirchoff theorem, allows to find out the geometrical
meaning of fermionic degrees of
freedom and of their boundary conditions (section two).
The continuum limit of dense polymers is discussed in section three. They
 are described by an $\eta,\xi$
system, whose pathologies as a conformal field theory are compatible
with the nature of the dense phase. For instance that $\xi$ has vanishing
 conformal
weight is associated with the existence of the density operator with non
vanishing expectation value. Partition functions are analyzed.
The $L$ legs operators are classified in three
sectors, Ramond, Neveu Schwartz, and a $Z_{4}$ sector for $L$ odd. Their
operator algebra is determined, as well as some four point functions.

The structure of sectors and the geometrical interpretation of fermions
 boundary conditions  holds also
for dilute polymers, where the $\eta,\xi$ system
has to be replaced by a twisted $N=2$ theory with $k=1$, whose degrees
 of freedom are
$\eta,\xi$ and two bosonic fields $\phi,\tilde{\phi}$. All results for dense
polymers are extended to the dilute case in section four.

Section five contains a discussion of the percolation problem, which can be
 mapped
onto  polymers at the theta point, the low
temperature Ising model, or to some extent brownian motion. It turns out to be
also described by twisted $N=2$ with $k=1$, which explains the coincidence of
percolation and dilute polymers exponents . A discussion of the same nature as
before is provided.

The conclusion contains some comments on the topological nature of some sectors
of our theory, and the geometrical meaning of the BRS operator in twisted
$N=2$.

As an application of the $N=2$ formalism, we discuss in the first appendix the
problem of percolation backbone, and predict in particular the exact valued of
the fractal dimension of the backbone to be $D=25/16$. In the second
 appendix we show how the introduction of fermions allows also
to give a meaning to the $n\rightarrow 0$ limit for the
off critical $O(n)$ model.

\section{Dense Polymers: Lattice Results}

\subsection{Dense polymers and Jordan Curves}
Recall that dense polymers are obtained by considering a finite number of self
avoiding and mutually avoiding loops or open chains that cover a finite
fraction of the available volume \cite{SA87,DS87}.
 They are known \cite{DS87,DD88} to possess some partially
solvable
realizations. Consider for instance a square lattice ${\cal L}$ and its
 {\bf medial} ${\cal M}$,
that is the other square lattice obtained by putting a vertex on every edge of
${\cal L}$, and joigning the vertices of ${\cal M}$ by an edge if they belong
to
perpendicular edges of ${\cal L}$ with a common end point. Consider now a {\bf
p
   olygon
decomposition} of ${\cal M}$ obtained by splitting  each vertex of ${\cal M}$
 in one of the two
possible ways shown in figure 2
and require that when connecting all the arcs only {\bf one} connected
 polygon is
obtained (this is usually called a {\bf Jordan Curve}). This polygon is
self avoiding and covers densely the medial lattice ${\cal M}$. Various
studies have confirmed that such a polygon has critical properties similar
to those of a dense polymer. Similarly if a finite number of polygons is
 formed by
connecting the arcs, the critical properties are those of several dense
polymers.\footnote{If a variable number of  polygons is allowed, and
controlled by a fugacity $\sqrt Q$, the critical properties are those of
the clusters boundaries in the $Q$ state Potts model \cite{SD87}}.

\subsection{Jordan Curves, Spanning Trees, and the Discrete Laplacian}

Suppose we color an edge of ${\cal L}$ every time the splitting of the
corresponding vertex of ${\cal M}$ does not intersect it, as illustrated in
figure 3.
Then it is easy to see that for every Jordan Curve, the colored edges draw a
{\bf Spanning Tree}, that is a subgraph of ${\cal L}$ without loops that
contain all its vertices. The number of spanning trees of any graph ${\cal G}$
is given by Kirchoff's theorem:

\smallskip

{\bf Theorem}(Kirchoff):

\smallskip

For any graph ${\cal G}$ with $V$ vertices denoted $i,j,\ldots$ define the
 discrete Laplacian $-\Delta$ to be the $V^{\otimes 2}$ matrix with entries
$-\Delta_{ii}$=number of edges incident to $i$,
 $-\Delta_{ij}$= - number of  edges with end points
$i$ and $j$. Then every principal minor of $-\Delta$ equals the number of
spanning trees of ${\cal G}$.

\bigskip

This result  leads for instance to the computation of
 the exponent gamma for dense
polymers \cite{DS87}. We are {\bf not} going to use it as it stands.
 The reason is
that the determinant of $\Delta$ vanishes, and to get a non
trivial information one has to
suppress the zero mode \footnote{In the $Q$ state Potts model
(resp. the $O(n)$ model in the dense phase) approach one takes a
$Q(\mbox{resp.}n)$ derivative at $Q(\mbox{resp.}n)=0$}: this is precisely
 what we want
 to avoid here.
We shall instead obtain results without taking any limit but using fermions.

\subsection{XX Chain}

Besides the discrete Laplacian, which is the  Lagrangian aspect of the
problem, it is also convenient to keep in mind the
 Hamiltonian point of view. This can
easily be done by realizing the above polygon decomposition algebraically in
the {\bf Temperley Lieb algebra} with relations
\begin{equation}
e_{i}^{2}=\delta e_{i}
\end{equation}
\begin{equation}
e_{i}e_{i\pm 1}e_{i}=e_{i}
\end{equation}
\begin{equation}
\left[e_{i},e_{j}\right]=0,\ |i-j|>1
\end{equation}
that can be realized graphically , ie by acting on polymers, as
 in figure 4 \cite{L87}. In our case we have to take $\delta=0$.

Corresponding to the above conventions the transfer matrix takes the form (for
a general system with anisotropy parameter $x$)
\begin{equation}
\tau=\prod_{i=1}^{{\cal L}}\left(1+xe_{2i}\right)
\prod_{i=1}^{{\cal L}-1}\left(x+e_{2i-1}\right)
\end{equation}
Instead of the loop realization we can use the vertex model realization. $\tau$
 describes now a {\bf 6 vertex} model on the square lattice with
a copy of $C^{2}$ per edge, diagonal propagation, and
\begin{equation}
e_{i}=1\otimes 1\ldots\otimes 1\otimes\left(\begin{array}{cccc}
0&0&0&0\\
0&-i&-1&0\\
0&-1&i&0\\
0&0&0&0
\end{array}\right)\otimes 1\ldots
\end{equation}
or
\begin{equation}
e_{i}=-\frac{1}{2}\left[\sigma^{x}_{i}\sigma^{x}_{i+1}+\sigma^{y}_{i}
\sigma^{y}_{i+1}+i(\sigma^{z}_{i}-\sigma^{z}_{i+1})\right]
\end{equation}
where $\sigma$'s are Pauli matrices. This  is the {\bf XX} hamiltonian
 with some boundary term. The role of the later
is to provide a {\bf zero mode}\cite{S89}. Introducing
\begin{equation}
V_{\pm}=
\sum_{j=1}^{{\cal L}}\prod_{k<j}i^{\sigma^{z}_{k}}\sigma^{\pm}_{j}
\end{equation}
one has
\begin{equation}
\left[e_{i},V_{\pm}\right]=0,\ \mbox{ for any i}
\end{equation}
while
\begin{equation}
\left(V_{\pm}\right)^{2}=0,\left\{V_{+},V_{-}\right\}=0
\end{equation}
The matrices $e_{i}$ commute also with the fermion number operator
\begin{equation}
 F=\sum_{j=1}^{L}\frac{\sigma_{j}^{z}+1}{2}
\end{equation}
such that
\begin{equation}
\left[F,V_{+}\right]=V_{+},\left[F,V_{-}\right]=-V_{-}
\end{equation}
This symmetry is $gl(1,1)/u(1)_{E}$. The casimir operator is simply
$V_{+}V_{-}$.

 Due to the presence of the zero mode, all levels of $\tau$ are
twice
degenerate, and occur once with odd fermion number, once with even fermion
number. Therefore
\begin{equation}
\mbox{Tr}\left[(-)^{F}\tau^{{\cal T}}\right]=0,\ \mbox{for any }{\cal T}
\end{equation}
This trace corresponds to computing the determinant of the discrete Laplacian
on a cylinder of size ${\cal T}\times {\cal L}$, that is with
 {\bf periodic} boundary
conditions in the time direction. In this case it is easy to see
 that the
contribution of non contractible loops vanishes because they are counted once
with a factor of $+1$, and once with a factor of $-1$. Therefore instead of
suppressing the zero mode "by hand" or by taking some appropriate limit, one
can get non trivial partition functions, and hence information about the dense
polymers, by changing the boundary conditions. Let us discuss this in more
details.

\subsection{Frustrating Some Edges}
We first establish the result

\smallskip

{\bf Lemma 2}:

\smallskip

Consider a graph\footnote{which we suppose without loss of
generality to have no multiple edges} ${\cal G}$ and some marked edge $[i|j]$.
 Consider the modified
discrete Laplacian $-\Delta[i|j]$ obtained  by changing from $-1$  to $+1$
the $ij$ and $ji$ matrix elements of $-\Delta$. The edge $[i|j]$ is said
to be {\bf frustrated}. One has then:

\smallskip

$\mbox{det}\left(-\Delta[i|j]\right)=4\times$ number of subgraphs of ${\cal
G}$ containing all its vertices
and made of {\bf one} loop passing
 through $[i|j]$ with dangling arms (trees) attached to it (see figure 5).

\smallskip

{\bf Proof}:

The proof of this statement follows closely the proof of Kirchoff's theorem.
Suppose the graph ${\cal G}$ has $V$ vertices and $E$ edges ($E\geq V-1$).
Give an arbitrary orientation to every edge and introduce the matrix $X_{\alpha
i}$ of size $E\times V$ where $\alpha$ runs over the oriented edges and $i$
 over the
vertices, with $X_{\alpha i}=\pm 1$ if the edge $\alpha$ has origin
 (extremity) in $i$ and  $X_{\alpha i}=0$ if the vertex $i$ is not one
 extremity of the edge
$\alpha$. Similarly introduce the matrix $Y$ deduced from $X$ by setting
$Y_{[i|j],i}=Y_{[i|j],j}=1$. It is then easy to check that
\begin{equation}
-\Delta=X^{t}X,\ -\Delta[i|j]=Y^{t}Y
\end{equation}
The modification of the discrete Laplacian suppresses the zero mode
and $\mbox{det}\left(-\Delta[i|j]\right)\neq0$. We can now write
\begin{equation}
\mbox{det}\left(Y^{t}Y\right)=\sum_{\alpha_{1}<\ldots<\alpha_{V}}\mbox{det}
\left(Y^{t}_{\alpha_{1}\ldots\alpha_{V}}
Y_{\alpha_{1}\ldots\alpha_{V}}\right)\label{eq:detexp}
\end{equation}
where $Y_{\alpha_{1}\ldots\alpha_{V}}$ is the square matrix obtained from $Y$
by keeping only the lines $\alpha_{1}\ldots\alpha_{V}$. The corresponding edges
draw a subgraph of ${\cal G}$, ${\cal G}_{\alpha_{1}\ldots\alpha_{V}}$.
Of course
 $\mbox{det}\left(Y_{\alpha_{1}\ldots\alpha_{V}}\right)$ is non zero if and
 only if
every vertex of ${\cal G}$ is in ${\cal G}_{\alpha_{1}\ldots\alpha_{V}}$
. It is also easy to see that
$\mbox{det}\left(Y_{\alpha_{1}\ldots\alpha_{V}}\right)$ vanishes unless the
 marked
edge $[i|j]$ is one of the labels $\alpha_{1}\ldots\alpha_{V}$, since it is the
modification of the $\alpha=[i|j]$ line that makes $Y$ of rank $V$, while $X$
is of rank $V-1$. Using Euler's relation
one finds that the subgraphs ${\cal G}_{\alpha_{1}\ldots\alpha_{V}}$
contributing to the summation (\ref{eq:detexp})
have  as many loops as components.
 Expand now $\mbox{det}\left(Y_{\alpha_{1}\ldots\alpha_{V}}\right)$
 along the $[i|j]$
line. This line has two non trivial entries that are equal to $1$, while
 for the matrix $X$ it contains a one
and a minus one. In the latter case
 $\mbox{det}\left(X_{\alpha_{1}\ldots\alpha_{V}}\right)$
vanishes because $X$ is of rank $V-1$. Therefore the two corresponding minors
have to be equal. Suppose first we are dealing with a one component subgraph,
and hence one loop. In the case where the loop does not pass through the marked
edge, the subgraphs associated with these minors contain one loop. Since these
minors are the same as the ones of $X$, they must vanish. In the case where the
loop does pass through the marked edge, the subgraphs associated with these
minors are trees, and they take value $\pm 1$. The corresponding
$\mbox{det}\left(Y_{\alpha_{1}\ldots\alpha_{V}}\right)$ equals then $\pm 2$.
 If we were
 dealing with several
components subgraphs, and hence several loops, the minors obviously would
vanish in any case.
Using  (\ref{eq:detexp}) the result is therefore established.

\smallskip

{\bf Example:}

Let us consider the  graph in figure 6a with some choice of edges orientation
We have
\[
L=\left(\begin{array}{cccc}
1&-1&0&0\\
0&1&-1&0\\
0&-1&0&1\\
-1&0&0&1\\
0&0&1&1
\end{array}\right);\ L^{t}=\left(\begin{array}{ccccc}
1&0&0&-1&0\\
-1&1&-1&0&0\\
0&-1&0&0&1\\
0&0&1&1&1
\end{array}\right)
\]
and therefore
\[
-\Delta[3|4]=\left(\begin{array}{cccc}
2&-1&0&-1\\
-1&3&-1&-1\\
0&-1&2&1\\
-1&-1&1&3
\end{array}\right)
\]
such that
\[
\mbox{det}\left(-\Delta[3|4]\right)=12
\]
corresponding to the three  subgraphs shown in figure 6b.

We shall use also the generalization of Lemma 2 to the case of several marked
edges:

\smallskip

{\bf Lemma 3:}

\smallskip

Consider a graph ${\cal G}$ and a set of marked edges
$[i_{1}|j_{1}],\ldots,[i_{k}|j_{k}]$. Then one has:

\smallskip

$\mbox{det}\left(-\Delta[i_{1}|j_{1}],\ldots,[i_{k}|j_{k}]\right)=$
sum over
subgraphs of ${\cal G}$ containing all its vertices
, that are
made of several loops
passing through an {\bf odd} number of marked edges
, and attached dangling
arms (that may also pass through some marked edges)
 of $4^{\mbox{number of components}}$

\smallskip

{\bf Proof:}

Let us start by a graph with two marked edges, say $[i|j]$ and $[k|l]$. Let us
introduce the matrix $Y$ and expand
 $\mbox{det}\left(-\Delta[i|j][k|l]\right)$ as above (\ref{eq:detexp}). The
only
subgraphs that may contribute are those where each vertex is extremity of at
least one of the edges $\alpha_{1},\ldots,\alpha_{V}$, and where at least one
of the marked edges appears. Also by Euler's relation there are as many loops
as components in each subgraph. If only one marked edge appears, we get the
same submatrix $Y_{\alpha_{1},\ldots\alpha_{V}}$ as above, hence the same
result about allowed subgraphs. Suppose now the two marked edges appear. If
they belong to two separate components, we can repeat the argument for each
separately. If they appear in the same component and none of them belongs to
the loop, the contribution vanishes. If one of them only belongs to the
loop the determinant is still equal to $\pm 2$. If both of them belong
 to the loop, let us expand
$\mbox{det}\left(Y_{\alpha_{1}\ldots\alpha_{V}}\right)$
 with respect to say the $[i|j]$
line. This line contains two non vanishing entries, equal to one. If one of
these entries was minus one instead we would be dealing with the conditions of
the preceding
theorem ,
and the determinant would be equal to $\pm 2$. Now opening $[i|j]$ edge leaves
a tree, that is the two minors of these two non vanishing entries have to be
themshelves equal to $\pm 1$. Therefore they are in fact opposite, and
$\mbox{det}\left(Y_{\alpha_{1}\ldots\alpha_{V}}\right)$ vanishes. Suppose
 now there is a
larger number of marked edges, and we consider some subgraph of ${\cal G}$ with
one loop that passes through n of them. Suppose we have proven that for $m$ odd
the resulting determinant is $\pm 2$, and for $m$ even it vanishes, $m<n$.
 Then expand
with respect to say the $[i|j]$ line. This line as above contains two ones. If
it contained a one and a minus one we would be dealing with a loop with $n-1$
marked edges. Hence by the induction assumption if $n-1$ is even the
determinant would vanish, if $n-1$ is odd the determinant would equal $\pm 2$.
Opening the $[i|j]$ edge leaves a tree, therefore the two minors are equal to
$\pm 1$. Hence if $n-1$ is even they take equal values, if $n-1$ is odd they
take opposite values. Therefore
 $\mbox{det}\left(Y_{\alpha_{1}\ldots\alpha_{V}}\right)$
equal $\pm 2$ if $n$ is odd and vanishes if $n$ is even. This establishes the
result.

\smallskip

{\bf Example:}

Consider the graph of the preceding example where we mark the third and fifth
edge. Therefore
\[
Y=\left(\begin{array}{cccc}
1&-1&0&0\\
0&1&-1&0\\
0&1&0&1\\
-1&0&0&1\\
0&0&1&1
\end{array}\right)
\]
One has
\[
-\Delta[2|4][3|4]=\left(\begin{array}{cccc}
2&-1&0&-1\\
-1&3&-1&1\\
0&-1&2&1\\
-1&1&1&3
\end{array}\right)
\]
with determinant equal to twelve. We give in the following some of the
determinants of the $Y_{\alpha_{1}\ldots\alpha_{V}}$
corresponding to the subgraphs of figure 7
\[
a)\mbox{det}\left(\begin{array}{cccc}
1&-1&0&0\\
0&1&-1&0\\
0&1&0&1\\
0&0&1&1
\end{array}\right)=0,\ b)
\mbox{det}\left(\begin{array}{cccc}
1&-1&0&0\\
0&1&-1&0\\
0&1&0&1\\
-1&0&0&1
\end{array}\right)=2
\]
\[
c)\mbox{det}\left(\begin{array}{cccc}
1&-1&0&0\\
0&1&0&1\\
-1&0&0&1\\
0&0&1&1
\end{array}\right)=-2,\ d)
\mbox{det}\left(\begin{array}{cccc}
1&-1&0&0\\
0&1&-1&0\\
-1&0&0&1\\
0&0&1&1
\end{array}\right)=-2
\]

\subsection{Non Contractible Polymers and the Laplacian
 with Antiperiodic Boundary
Conditions}

As an application of this last lemma, consider a rectangle of the square
lattice with doubly periodic boundary conditions, that is wrapped on a torus.
Suppose we frustrate all the horizontal edges along a given column. We refer to
this column as a {\bf frustration line}. The
corresponding modified Laplacian is now what may be called the Laplacian with
antiperiodic boundary conditions in the time direction, and still periodic
boundary conditions in the space direction.
 We denote it by $-\Delta_{AP}$. Due
to the position of frustrated edges, the only loops that can occur in the
expansion of the corresponding determinant according to the rules of Lemma 3
are
{\bf non contractible}, and moreover must intersect the frustration line an
{\bf odd} number of times. They contribute then by a factor of $4$ to the
summation. Now consider
the medial lattice; to each such non contractible loop correspond two polymers
which are as well non contractible and intersect the frustration line an odd
number of times. This result immediately generalizes to the case where the
frustration line is along the time axis. We can therefore write

\smallskip

{\bf Lemma 4:}

\smallskip

$\mbox{det}\left(-\Delta_{AP(resp.PA)}\right)=$ Sum over dense
 coverings of the
medial
by an {\bf even} number of (non contractible) polymers that cross
$\omega_{2}(\mbox{resp.}\omega_{1})$
an
{\bf odd} number of times of  $2^{\mbox{number of polymers}}$

\bigskip

while if we put a pair of frustration lines along space and time axis, it is
the total number of intersections with these axis that has to be odd:

\smallskip

{\bf Lemma 5:}

\smallskip

$\left(\mbox{det}-\Delta_{AA}\right)=$ Sum over dense
 coverings of the
medial
by an {\bf even} number of (non contractible) polymers that cross
$(\omega_{1},\omega_{2})$
an
{\bf odd} number of times of $ 2^{\mbox{number of polymers}}$

\bigskip

Let us recall also that
\begin{equation}
\mbox{det}\left(-\Delta_{PP}\right)=0
\end{equation}
\[
\mbox{det'}\left(-\Delta_{PP}\right)=\left\{\mbox{number of configurations
 of a {\bf
single}}\right.
\]
\begin{equation}
\left.\mbox{{\bf contractible} polymer densely covering the medial}\right\}
\end{equation}
As we remarked above, with periodic boundary conditions the contribution of
contractible loops disappears because they are counted once with a weight $1$,
once with $-1$. This corresponds to the fact that they can be described either
by a boson or by a fermion, with opposite contributions due to statistics
. If we now impose {\bf antiperiodic} boundary conditions for the
fermion, the weight of a non contractible polymer becomes
\begin{equation}
1+(-1)^{(\mbox{number of intersections with the frustration line $+1$})}
\end{equation}
thereby recovering the result of the above two lemmas.

\section{The Continuum Limit of Dense Polymers}

\subsection{Continuum Limit Partition Functions of Dense Polymers}

\subsubsection{$Z_{2}$ Sector}
Let us define the lattice partition functions of dense polymers by the
expressions of lemmas $4$ and $5$ that is
\begin{eqnarray}
{\cal Z}_{AP}=\mbox{det}\left(-\Delta_{AP}\right)\nonumber\\
{\cal Z}_{PA}=\mbox{det}\left(-\Delta_{PA}\right)\nonumber\\
{\cal Z}_{AA}=\mbox{det}\left(-\Delta_{AA}\right)
\end{eqnarray}
The continuum limit of these partition functions is obtained using standard
free field computations. In general for the continuum Laplacian $D$ with
twisted boundary conditions characterized by the phases $exp(2i\pi k/N)$ along
$\omega_{1}$, $exp(2i\pi  l/N)$ along $\omega_{2}$ one has (see
e.g.\cite{IS87})
\begin{equation}
\mbox{det}\left(-D_{l/N,k/N}\right)=|d_{l/N,k/N}|^{2}
\end{equation}
where
\begin{equation}
d_{l/N,k/N}=q^{\left[1-6k/N(1-k/N)\right]/12}\prod_{n=0}^{\infty}
\left(1-e^{2i\pi l/N}q^{n+k/N}\right)\prod_{n=0}^{\infty}
\left(1-e^{-2i\pi l/N}q^{n+1-k/N}\right)
\end{equation}
Therefore
\begin{eqnarray}
Z_{AP}=\left|d_{1/2,0}\right|^{2}=
4\left|q^{1/12}\prod_{n=1}^{\infty}\left(1+q^{n}
\right)^{2}\right|^{2}\nonumber\\
Z_{PA}=\left|d_{0,1/2}\right|^{2}=
\left|q^{-1/24}\prod_{n=0}^{\infty}\left(1-q^{n+1/2}
\right)^{2}\right|^{2}\nonumber\\
Z_{AA}=\left|d_{1/2,1/2}\right|^{2}=\left|q^{-1/24}\prod_{n=0}^{\infty}\left(
1+q^{n+1/2}\right)^{2}\right|^{2}\label{eq:detprod}
\end{eqnarray}
Define now the manifestly modular invariant quantity
\[
{\cal Z}^{e}=\left\{\mbox{Sum over dense coverings of the medial by an even
 number}\right.
\]
\begin{equation}
\left.\mbox{of non contractible polymers with weight  } 2^{ \mbox{number of
 polymers}}\right\}
\end{equation}
One checks easily that
\begin{equation}
{\cal Z}^{e}=\frac{1}{2}\left({\cal Z}_{AP}+{\cal Z}_{PA}+{\cal Z}_{AA}\right)
\end{equation}
and therefore in the continuum limit
\begin{equation}
Z^{e}=\frac{1}{2}\left[\mbox{det}\left(-D_{1/2,0}\right)+
\mbox{det}\left(-D_{0,1/2}\right)+\mbox{det}\left(-D_{1/2,1/2}\right)\right]
\end{equation}
or
\begin{equation}
Z^{e}=\frac{1}{2}\left[\left|d_{0,1/2}\right|^{2}+\left|d_{1/2,0}\right|^{2}+
\left|d_{1/2,1/2}\right|^{2}\right]
\end{equation}
That the partition function with doubly periodic boundary conditions vanishes
 is characteristic of the dense phase. Although a
naive examination of the loop expansion of say the $O(n)$ model gives a result
one for this partition function (because all polymers disappear), as soon
 as some polymer is allowed by changing the boundary conditions, the partition
function of the system grows {\bf exponentially} with the volume due to the
finite density. We thus have
\begin{equation}
{\cal Z}_{AP}\propto{\cal Z}_{PA}\propto{\cal Z}_{AA}\propto
e^{f\times\mbox{area}},\ {\cal Z}_{PP}\propto 1\label{eq:freeenergy}
\end{equation}
 When we write continuum limit partition functions we always
discard such factors because they are non universal and generally do not depend
on boundary conditions (this is not the case is this broken symmetry phase, and
for instance $n\rightarrow 0$ limit does {\bf not} commute with thermodynamic
limit or changing the boundary conditions). For consistency we therefore have
to
set $Z_{PP}=0$.

\subsubsection{Comparison with Lattice Coulomb Gas Computations}

Now the mapping of dense polymers, that is the low temperature phase of the
$O(n)$ model for $n\rightarrow 0$, on a Coulomb Gas \footnote{In the following
w
   e
refer to the mapping of the $O(n)$ model or $Q$ state Potts model onto a free
field as lattice Coulomb gas. This is related, but not strictly equivalent to
the usual Dotsenko Fateev representation holding for continuum theories}
, allows another evaluation
of the above partition functions. We will not recall here how this works but
merely state some results that can easily be extracted from \cite{DFSZ87,DS87}.
Introduce the coupling constant $g$ related to $n$ by
\begin{equation}
n=-2cos\pi g,\ g\in[0,1]\mbox{(dense)},\ g\in[1,2]\mbox{(dilute)}
\end{equation}
hence
\begin{equation}
g=\frac{1}{2}
\end{equation}
in the present case. Introduce also
\begin{equation}
Z_{mm'}(g)=\sqrt{\frac{g}{\mbox{Im}\tau}}\frac{1}{\eta\overline{\eta}}
\mbox{exp}\left(\frac{-\pi g}{\mbox{Im}\tau}
\left|m-m'\tau\right|^{2}\right)
\end{equation}
where $\eta (q)$ is the Dedekind function (not to be confused with the fermions
introduced in the next section)
\begin{equation}
\eta=q^{1/24}\prod_{n=1}^{\infty}\left(1-q^{n}\right),\ q=e^{2i\pi \tau}
\end{equation}
Then one finds, using the above definitions in terms of non contractible loops
and translating them in the Coulomb Gas language\footnote{The normalization of
these continuum limit expressions is not fixed by the Coulomb gas mapping, and
chosen for future convenience}
\begin{eqnarray}
Z_{PP}=\frac{1}{2}\sum_{mm'\in Z}(-)^{m\wedge m'}Z_{mm'}\nonumber\\
Z_{AP}=\frac{1}{2}\sum_{mm'\in Z}(-)^{m}(-)^{m\wedge m'}Z_{mm'}\nonumber\\
Z_{PA}=\frac{1}{2}\sum_{mm'\in Z}(-)^{m'}(-)^{m\wedge m'}Z_{mm'}\nonumber\\
Z_{AA}=\frac{1}{2}\sum_{mm'\in Z}(-)^{m+m'}(-)^{m\wedge m'}Z_{mm'}
\end{eqnarray}
where $m\wedge m'$ means the greatest common divisor of $m$ and $m'$. These
expressions become after Poisson resummation
\[
Z_{PP}=\frac{1}{2\eta\overline{\eta}}\left(\sum_{m\in 2Z,e\in
Z/2}q^{h_{em}}\overline{q}^{\overline{h}_{em}}-\sum_{m\in Z,e\in
Z}q^{h_{em}}\overline{q}^{\overline{h}_{em}}\right)
\]
\[
Z_{AP}=\frac{1}{\eta\overline{\eta}}\sum_{m\in 2Z,e\in
Z+1/2}q^{h_{em}}\overline{q}^{\overline{h}_{em}}
\]
\[
Z_{PA}=\frac{1}{2\eta\overline{\eta}}\left(\sum_{m\in 2Z,e\in Z}-\sum_{m\in
2Z+1,e\in Z+1/2}\right)q^{h_{em}}\overline{q}^{\overline{h}_{em}}
\]
\begin{equation}
Z_{AA}=\frac{1}{2\eta\overline{\eta}}\left(\sum_{m\in 2Z,e\in Z}+\sum_{m\in
2Z+1,e\in Z+1/2}\right)q^{h_{em}}\overline{q}^
{\overline{h}_{em}}\label{eq:detsum}
\end{equation}
where
\begin{equation}
h_{em}=\frac{1}{4g}\left(e+mg\right)^{2},\ \overline{h}_{em}=\frac{1}{4g}\left(
e-mg\right)^{2}
\end{equation}
Equality of the expressions (\ref{eq:detsum}) and (\ref{eq:detprod}) is
 readily checked using
Jacobi identity. One finds also
\begin{equation}
Z^{e}=\frac{1}{2\eta\overline{\eta}}\sum_{m\in 2Z,e\in Z/2}
q^{h_{em}}\overline{q}^{\overline{h}_{em}}=\frac{1}{2}Z_{c}\left[g=1/2,f=2
\right]
\end{equation}
(A similar identity occurs when bosonizing a free complex fermion)
where we defined the Coulomb partition functions
\begin{equation}
Z_{c}\left[g,f\right]=\frac{1}{\eta\overline{\eta}}\sum_{m\in fZ,e\in Z/f}
q^{h_{em}}\overline{q}^{\overline{h}_{em}}
\end{equation}
Hence Coulomb Gas computations are in complete agreement with the lattice
analysis.

 It must be stressed that the dense phase of polymers has critical
properties independent of the density $\rho>0$.
 Therefore the above expressions are
expected to hold in general, while the lattice analysis based on the Laplacian
occurs in the maximum density limit.

Now, from the general point of view of dense polymers it is natural to define
still another partition function
\[
{\cal Z}^{o}=\left\{\mbox{Sum over dense coverings of the lattice by an
 {\bf odd}
 number}\right.
\]
\begin{equation}
\left.\mbox{of non contractible polymers of } 2^{ \mbox{number of polymers}}
\right\}
\end{equation}
This object does not appear in consideration of Jordan Curves and the discrete
Laplacian. Its continuum limit has been computed in the Coulomb Gas mapping
\begin{equation}
Z^{o}=Z_{c}\left[1/2,1/2\right]-\frac{1}{2}Z_{c}\left[1/2,2\right]
\end{equation}
It is useful to recall that
\begin{equation}
Z_{c}\left[1/2,1\right]=Z_{c}\left[1/2,2\right]\label{eq:relation}
\end{equation}
Finally the total dense polymers partition function turns out to be
\begin{equation}
{\cal Z}^{e+o}={\cal Z}^{e}+{\cal Z}^{o}\rightarrow\ Z^{e+o}=
Z_{c}\left[1/2,1/2\right]
\end{equation}
where the arrow indicates the continuum limit. Notice that the ground state
with $h=0$ is {\bf twice} degenerate. This is expected in the $\eta,\xi$ system
where both the identity and the $\xi$ field have vanishing conformal
weight. This is also
natural from the polymer point of view since, besides the identity, the density
operator $\rho$ has a non vanishing expectation value \footnote{We
 remind the reader that the density is however not
 the order parameter from the point of view of the $O(n)$ model} and therefore
must have $h=0$.

\subsubsection{$Z_{4}$ Sector}

One of the major obstacles in building the consistent theory of dense polymers
has been the presence of the  sector with an odd number of polymers that has
 no obvious lattice interpretation
in terms of the discrete Laplacian. However {\bf in the continuum} it seems a
reasonable guess to expect that the description of the odd sector involves
$Z_{4}$ twists\footnote{There are actually some lattice arguments to justify
this, based on considering the XX chain on a circle, with twisted boundary
conditions}. It is an elementary exercise in applying the Jacobi identity to
check the following nice formula
\begin{eqnarray}
Z^{o}=\frac{1}{2}\left[\mbox{det}\left(-D_{0,1/4}\right)+
\mbox{det}\left(-D_{1/4,0}\right)+\mbox{det}
\left(-D_{1/4,1/4}\right)\right.\nonumber\\
\left.+\mbox{det}\left(-D_{1/2,1/4}\right)+\mbox{det}\left(-D_{1/4,1/2}\right)
+\mbox{det}\left(-D_{3/4,1/4}\right)\right]
\end{eqnarray}
or
\begin{equation}
Z^{o}=\frac{1}{2}\left[\left|d_{0,1/4}\right|^{2}+\left|d_{1/4,0}\right|^{2}+
\left|d_{1/4,1/4}\right|^{2}+\left|d_{1/2,1/4}\right|^{2}+
\left|d_{1/4,1/2}\right|^{2}+\left|d_{3/4,1/4}\right|^{2}\right]
\end{equation}
Having identified the correct object related to the Laplacian that reproduces
$Z^{o}$ we shall {\bf assume} that this is not a numerical coincidence but the
correct answer in the continuum limit. We can now proceed and study the
corresponding conformal field theory.

\subsection{The Conformal Field Theory of Dense Polymers}

\subsubsection{Generalities}

To reproduce determinants of the Laplacian with various boundary conditions we
introduce an ${\bf \eta,\xi}$ system with action
\begin{equation}
S=\frac{1}{\pi}\int d^{2}z
\left(\eta\overline{\partial}\xi+\overline{\eta}\partial\overline{\xi}\right)
\end{equation}
where $\eta$ and $\xi$ denote conjugate {\bf Fermi}
fields of dimension $1$ and $0$
respectively. Short distance limit of the operator product is
\begin{equation}
\eta(z)\xi(w)\rightarrow\frac{1}{z-w},\ \xi(z)\eta(w)
\rightarrow\frac{1}{z-w}\ \mbox{as }z\rightarrow w
\label{eq:fermionspropagator}
\end{equation}
and the stress energy tensor reads
\begin{equation}
T(z)=-:\eta\partial\xi: \label{eq:stress}
\end{equation}
with central charge
\begin{equation}
c=-2
\end{equation}
This system is the free field representation for a current
algebra based on $gl(1,1)/u(1)_{E}$, the symmetry of dense polymers that was
 exhibited in the hamiltonian point of view (XX chain). We consider now
 the properties of arbitrary twist fields in the
$\eta,\xi$ system. The twist field $\sigma_{k/N}$
  can be defined via the follwing short
distance expansions ($k/N>0$)
\begin{eqnarray}
\eta(z)\sigma_{k/N}\left(w,\overline{w}\right)\propto(z-w)^{-k/N}\tau_{k/N}\\
\partial\xi(z)\sigma_{k/N}\left(w,\overline{w}\right)\propto(z-w)^{-1+k/N}
\tau_{k/N}'\\
\overline{\eta}\left(\overline{z}\right)\sigma_{k/N}\left(w,\overline{w}\right)
\propto\left(\overline{z}-\overline{w}\right)^{-1+k/N}\tilde{\tau}_{k/N}\\
\overline{\partial}\ \overline{\xi}\left(\overline{z}\right)\sigma_{k/N}
\left(w,\overline{w}\right)
\propto\left(\overline{z}-\overline{w}\right)^{-k/N}
\tilde{\tau}_{k/N}'\label{eq:mono}
\end{eqnarray}
In this formulas the non integer powers of the singularities are determined by
the condition that $\eta$ or $\xi$ field picks up a phase $e^{2i\pi k/N}$ when
rotated around the origin (while the stress tensor has to remain periodic).
 The $\tau$'s are excited twist fields. The primes
distinguish between two different types of excited twist fields with different
values of $h$ but the same $\overline{h}$ or vice versa. The tilde denoted
fields related by complex conjugation $h\leftrightarrow\overline{h}$,
$z\leftrightarrow\overline{z}$. The integer powers are determined as follows.
Let us write the mode expansion
\begin{eqnarray}
\eta(z)=\sum_{m=-\infty}^{\infty}\eta_{m+k/N}z^{-m-1-k/N}\\
\xi(z)=\sum_{m=-\infty}^{\infty}\xi_{m-k/N}z^{-m+k/N}
\end{eqnarray}
where, from the short distance singularity (\ref{eq:fermionspropagator})
 one has the
anticommutation relation
\begin{equation}
\left\{\xi_{m+k/N},\eta_{n-k/N}\right\}=\delta_{m+n}
\end{equation}
while all other anticommutators vanish.
The ground state of the $k/N$ twisted sector is annihilated by all the positive
frequency mode operators $\eta_{m-1+k/N},\xi_{m-k/N}$, $m>0$. The excited twist
fields in the expression (\ref{eq:mono}) can be expressed as descendents of
$\left|\sigma_{k/N}\right>$ like for instance
$\left|\tau_{k/N}\right>$ $=$ $\eta_{-1+k/N}\left|\sigma_{k/N}\right>$.
The conformal weight of the operator $\sigma_{k/N}$ is easily computed
following the technique of \cite{DFMS87}. One introduces the object
\begin{equation}
g(z,w)=-\frac{\left<\eta(z)\partial\xi(w)\sigma_{+}(z_{1})\sigma_{-}(z_{2})
\right>}{\left<\sigma_{+}(z_{1})\sigma_{-}(z_{2})\right>}
\end{equation}
where $\sigma_{+}$ stands for $\sigma_{k/N}$, $\sigma_{-}$ stands for its
"anti twist", whose definition is similar to (\ref{eq:mono}) but with
 $k/N$ replaced
by $1-k/N$. $g$ is uniquely determined by the various monodromy constraint,
while letting $z$ go to $w$ allows to extract the stress energy tensor, and
hence the required weight. One finds
\begin{equation}
h=-\frac{1}{2}\frac{k}{N}\left(1-\frac{k}{N}\right)\label{eq:weight}
\end{equation}
Owing to the local relation (\ref{eq:stress}) we have
\[
L_{n}=\sum_{m=-\infty}^{\infty}\left(n-m-k/N\right)\eta_{m+k/N}\xi_{n-m-k/N},
\mbox{for }\  n\neq 0
\]
\begin{eqnarray}
L_{0}=-\frac{1}{2}\frac{k}{N}\left(1-\frac{k}{N}\right)
+\sum_{m=1}^{\infty}\left[\left(m-1+k/N\right)\xi_{-m+1-k/N}
\eta_{m-1+k/N}\right.\nonumber\\
\left.+\left(m-k/N\right)\eta_{-m+k/N}\xi_{m-k/N}\right]\label{eq:vir}
\end{eqnarray}

\subsubsection{Operator Content of The Various Sectors}

Let us start with the antiperiodic (Ramond) sector. One finds easily that
\begin{equation}
\left|d_{l/2,1/2}\right|^{2}=\frac{1}{\eta}\left(1+2
\sum_{n=1}^{\infty}(-1)^{(l+1)n}q^{n^{2}/2}\right)\times\mbox{cc}
\label{eq:detanti}
\end{equation}
We see that the ground state of the Ramond sector has conformal weight
\begin{equation}
h=-\frac{1}{8}
\end{equation}
in agreement with (\ref{eq:weight}). The physical meaning of
 the operator $\sigma_{1/2}$ in the polymer system is
the following. Suppose we have two punctures in the plane, and two frustration
lines going from the points to infinity. Following Lemma 3 the various
configurations that contribute to this object are made of loops encircling one
point, with dangling ends attached to them, counted with a factor of $4$. The
dominant contributions correspond to polymers
that separate the two points by encircling one of them, with a factor two
(figure 8). They
are selected in the Coulomb gas  by putting an
electric  charge
$-e_{0}$ ($e_{0}=1-g$) at both points, such that the polymer loops instead of
 having a weight
$e^{i\pi e_{0}}+e^{-i\pi e_{0}}=0$ get the weight $1+1=2$. The associated
dimension is $h=-e_{0}^{2}/4g=-1/8$ in agreement with the above result. The
$L_{0}$ generator in that sector writes
\begin{equation}
L_{0}=\frac{-1}{8}+\sum_{m=1}^{\infty}(m-1/2)\left(\xi_{-m+1/2}\eta_{m-1/2}+
\eta_{-m+1/2}\xi_{m-1/2}\right)\label{eq:ham}
\end{equation}
 In addition to the ground state
$|\sigma_{1/2}>$, other primary fields appear with conformal weights given by
\begin{equation}
h_{4l}=-\frac{1}{8}+\frac{l^{2}}{2}
\end{equation}
They all have even multiplicity (for $l>0$)
due to the $\eta,\xi$ symmetry in (\ref{eq:ham}). Their fermion
 number is equal to one (zero)
for $l$ odd ($l$ even). The first one has $h=3/8$ corresponding to
$\eta_{-1/2}\left|\sigma_{1/2}\right>$ and $\xi_{-1/2}
\left|\sigma_{1/2}\right>$. The weights can be
reproduced by the Kac formula and one finds
\begin{equation}
h_{4l}=h_{1,2+2l}\label{eq:weig}
\end{equation}
The exact decomposition of the sum (\ref{eq:detanti})is provided by
 expansion on $c=-2$ characters of the various terms. One has,
following the appendix of \cite{DFSZ87}
\begin{eqnarray}
\frac{1}{\eta}q^{2n^{2}}=\sum_{k=n}^{\infty}\chi_{1,2+4k}\nonumber\\
\frac{1}{\eta}q^{(2n+1)^{2}/2}=\sum_{k=n}^{\infty}
\chi_{1,4k+4}\label{eq:chaanti}
\end{eqnarray}
On the other hand,
from the lattice point of view, a natural
family of operators in the antiperiodic sector is obtained as follows. We
can imagine selecting the configurations where say $l$ of the loops pinch the
two punctures as in figure 9. This is a $4l$ legs operator, with dimension from
the Lattice Coulomb gas mapping given precisely by (\ref{eq:weig}).
As we observe these dimensions coincide also
with those appearing in the expansion
(\ref{eq:detanti}) as well as (\ref{eq:chaanti}). An important observation is
that the $L$ legs operators appear for $L=0\mbox{ mod }4$ in the Ramond sector.
This modulo $4$ occurs because the fundamental objects are trees rooted to some
loops, and on the medial graph these are associated to even number of polymer
loops.
 Such  modulo $4$ is also observed in the periodic (Neveu Schwartz)
sector to which we turn now.

We have
\begin{equation}
\left|d_{l/2,0}\right|^{2}=\frac{1}{\eta}\left(\sum_{n=-\infty}^{\infty}
(-)^{(l+1)n}q^{(2n+1)^{2}/8}\right)\times\mbox{cc}\label{eq:detperi}
\end{equation}
while we can write
\begin{equation}
L_{n}=-\sum_{-\infty}^{\infty}(m-n)\eta_{m}\xi_{n-m}
\end{equation}
\begin{equation}
L_{0}=\sum_{1}^{\infty}m\left(\xi_{-m}\eta_{m}+\eta_{-m}\xi_{m}\right)
\end{equation}
The $SL_{2}$ invariant vacuum is defined by $\eta_{n}|0>=0$, $n\geq 0$,
$\xi_{n}|0>=0$, $n>0$. The decomposition of (\ref{eq:detperi}) among
irreducible
representations of the
Virasoro algebra is accomplished via the formulas
\begin{eqnarray}
\frac{1}{\eta}q^{(4n+1)^{2}/8}=\sum_{k=n}^{\infty}\chi_{1,3+4k}+
\chi_{1,5+4k},\ n\geq 0\nonumber\\
\frac{1}{\eta}q^{(4n-1)^{2}/8}=\sum_{k=n}^{\infty}\chi_{1,1+4k}+
\chi_{1,3+4k},\ n>0
\end{eqnarray}
The ground state is {\bf twice} degenerate, as discussed above\footnote{However
only the identity provides an $SL_{2}$ invariant vacuum}.
 Due to the well known property that  $<0|\xi|0>\neq 0$ we identify the density
$\rho$ with $\xi+\overline{\xi}$.
 The other operators have conformal weights given by $h_{1,3+4k}$ and
$h_{1,5+4k}$ ($k\geq 0$). These coincide with the dimensions of the $4l+2$
 legs operators
\begin{equation}
h_{4l+2}=-\frac{1}{8}+\frac{(l+1/2)^{2}}{2}
\end{equation}

We turn finally to the $Z_{4}$ sector. We find
\begin{equation}
\left|d_{l/4,1/4}\right|^{2}=\frac{1}{\eta}\sum_{n=-\infty}^{\infty}
(-1)^{(1-l/2)n}q^{(4n+1)^{2}/32}\times\mbox{cc}\label{eq:detz4}
\end{equation}
The ground state $\left|\sigma_{1/4}\right>$ has conformal weight
\begin{equation}
h=-\frac{3}{32}
\end{equation}
In the polymer language this is the dimension of the one leg operator (its
negative value was interpreted and discussed in \cite{SA87}). In addition to
this other primary fields appear in the expansion with weights
\begin{eqnarray}
h_{4l+1}=-\frac{1}{8}+\frac{(l+1/4)^{2}}{2}\nonumber\\
h_{4l+3}=-\frac{1}{8}+\frac{(l+3/4)^{2}}{2}
\end{eqnarray}
They can be reproduced formally by the Kac formula with half integer labels
\begin{eqnarray}
h_{4l+1}=h_{1,2l+5/2}\nonumber\\
h_{4l+3}=h_{1,2l+7/2}
\end{eqnarray}
The corresponding fields are therefore not degenerate with respect to the
Virasoro algebra and the character expansion of (\ref{eq:detz4}) can be
easily written as
\begin{equation}
\left|d_{l/4,1/4}\right|^{2}=\sum_{n=0}^{\infty}
(-1)^{(1-l/2)n}\left(\chi_{1,2n+5/2}+\chi_{1,2n+7/2}\right)
\end{equation}
One has in particular that $\xi_{-1/4}\left|\sigma_{1/4}\right>$ is primary
with weight $h=5/32$, $\eta_{-3/4}\left|\sigma_{1/4}\right>$ is primary with
weight $h=21/32$. These operators are respectively the $3$ and $5$ legs
operators.

The content in polymers operators of the various sectors is therefore the
following
\begin{eqnarray}
\mbox{Neveu Schwartz}:\ L=2\mbox{ mod}\ 4\nonumber\\
\mbox{Ramond}:\ L=0\mbox{ mod}\ 4\nonumber\\
Z_{4}:\ L=1,3\mbox{ mod}\ 4
\end{eqnarray}
Moreover in each sector the fields are descendents of the one with lowest value
of $L$ with respect to the $\eta,\xi$ algebra. So there are only three
fundamental geometrical operators.
\subsection{$U(1)$ Current}

The natural current of the $\eta,\xi$ system is given by
\begin{equation}
J=-:\eta\xi:
\end{equation}
The associated $U(1)$ charge $J_{0}$ is denoted $Q$ in the following (we
 consider only left $U(1)$ charge, as we considered only the left moving sector
before). One
 has $Q=1$ for
$\xi$, $Q=-1$ for $\eta$. As was noticed in \cite{DFMS87}, when the vacuum is
defined as the $SL_{2}$ invariant state $|0>$, $J$ is not a primary field and
one has
\begin{equation}
T(z)J(w)\rightarrow \frac{-1}{(z-w)^{3}}+\frac{J(z)}{(z-w)^{2}}
\end{equation}
and hence
\begin{eqnarray}
\left[L_{1},J_{-1}\right]=J_{0}-1\nonumber\\
\left[L_{1},J_{-1}\right]^{\dagger}=\left[L_{-1},J_{1}\right]=-J_{0}
\end{eqnarray}
The system has therefore charge assymmetry, $Q^{\dagger}=1-Q$.
. For the
polymer density operator $\rho$ we associate the value $Q_{2}=1$ in such a way
that $<0|\rho|0>\neq 0$ holds. The other $4l+2$ legs operators in the Neveu
Schwartz sector therefore have charge $Q_{4l+2}=1+l$.
 The $U(1)$ charge of the twist fields $\sigma_{k/N}$ is $Q=k/N$. To the
 ground state of the Ramond sector we can associate the charge $Q=1/2$.
Therefore the $L=4l$ legs operators have $Q_{4l}=1/2+l$. To the ground state of
the $Z_{4}$ sector we associate the charge $Q=3/4$, in such a way that the
$L=4l+1$ operators have charge $Q_{4l+1}=3/4+l$, the $L=4l+3$ operators have
$Q_{4l+3}=5/4+l$. in summary for the $L$ legs operator charges  given by
\begin{equation}
\Phi_{L}\leftrightarrow Q=\frac{1}{2}\pm\frac{L}{4}
\end{equation}
Hence $Q$ counts essentially the number of legs, and one can think of $J$ as
describing the correlations of orientation along the polymer chains \cite{M90}.

\subsubsection{Correlation Functions and Operator Algebra}

 In polymer theory one is mainly interested in correlations of $L$ legs
operators. For $L$ even, these operators have dimensions in the Kac table.
Their four point functions can therefore be computed in the Coulomb Gas
formalism, that is by bosonizing the $\eta,\xi$ system and introducing
screening
charges. They can also be computed directly in the $\eta,\xi$ system.
The consistency of the two approaches is expected, but not always
 straightforward
to check. Let us consider first the Neveu Schwartz sector and
 the example of the $6$ legs operator that has
$h=\overline{h}=1$. These weights are reproduced by $h_{21}$ in the Kac table.
The four point function needs therefore introduction of a single screening
charge. One finds, using the formula (4.18) of \cite{DF84}
\begin{equation}
\left<\Phi_{6}(1)\Phi_{6}(2)\Phi_{6}(3)\Phi_{6}(4)\right>\propto
\frac{1}{\left|z_{13}\right|^{4}\left|z_{24}\right|^{4}}\frac{1}{\left|x(1-x)
\right|^{4}}\left[3\left|x^{3}(2-x)\right|^{2}+\left|2-4x+2x^{3}-x^{4}\right|
^{2}\right]
\end{equation}
where $x=z_{12}z_{34}/z_{13}z_{24}$. This can be recast as
\[
\left<\Phi_{6}(1)\Phi_{6}(2)\Phi_{6}(3)\Phi_{6}(4)\right>=
\]
\begin{eqnarray}
\frac{2}{\left|z_{13}\right|^{2}\left|z_{24}\right|^{2}}\frac{1}{\left|x(1-x)
\right|^{4}}\left\{2\left[x^{2}\overline{x}^{2}+(1-x)^{2}(1-\overline{x})^{2}
+x^{2}\overline{x}^{2}(1-x)^{2}(1-\overline{x})^{2}\right]\right.\nonumber\\
\left.-\left[x^{2}\overline{x}^{2}(1-x)^{2}+\overline{x}^{2}(1-x)^{2}
+x^{2}(1-x)^{2}
(1-\overline{x})^{2}+\mbox{cc}\right]\right\}\label{eq:}
\end{eqnarray}
This is the correlator of the
operator$\left[\eta(z)+\partial\xi(z)\right]\times\mbox{cc}$, computed easily
in the free fermion theory.

For the other sectors, we need the knowledge of the four point functions
 of twist operators. They are determined using the general
technique of \cite{DFMS87}. One introduces the auxiliary function
\begin{equation}
g\left(z,w;z_{i}\right)=-\frac{\left<\eta(z)\partial\xi(w)\sigma_{+}(1)
\sigma_{-}(2)\sigma_{+}(3)\sigma_{-}(4)\right>}{\left<\sigma_{+}(1)
\sigma_{-}(2)\sigma_{+}(3)\sigma_{-}(4)\right>}\label{eq:auxfct}
\end{equation}
which is determined by the local and global monodromy conditions. Letting $z$
go to $w$ allows to extract the stress tensor in (\ref{eq:}) and therefore, by
letting further $z$ go to one of the $z_{i}$'s, a linear differential equation
satisfied by the four point function, which can easily be integrated. One finds
\[
\left<\sigma_{+}(1)
\sigma_{-}(2)\sigma_{+}(3)\sigma_{-}(4)\right>\propto
\]
\begin{equation}
\left|
z_{12}z_{34}\right|^{2k/N(1-k/N)}\left|x(1-x)\right|^{2k/N(1-k/N)}
\left[F_{k/N}(x)\overline{F_{k/N}}(1-\overline{x})+F_{k/N}(1-x)
\overline{F_{k/N}}
(\overline{x})\right]\label{eq:twistcor}
\end{equation}
where $F$ is the hypergeometric function
\begin{equation}
F_{k/N}(x)=F(k/N,1-k/N;1,x)
\end{equation}
Consider for instance the ground state of the Ramond sector, ie
$k/N=1/2$. In that case twist and anti twist are identical. The conformal
weights are $h=\overline{h}=-1/8$ and coincide with the dimensions $h_{12}$ in
the Kac table. Again the four point function could also be computed in the
Coulomb Gas formalism. One screening charge is needed. The combination of
conformal blocks entering the result in formula (4.18) of \cite{DF84} turns out
however to vanish
identically. To get a finite result one has to consider the $c\rightarrow -2$
limit (i.e. again a $n\rightarrow 0$ limit!). One finds that
\[
\left<\Phi_{12}(1)\Phi_{12}(2)\Phi_{12}(3)\Phi_{12}(4)\right>\propto
\left|z_{13}z_{24}\right|^{1/2}\left|x(1-x)\right|^{1/2}
\left|F(1/2,1/2;1,x)\right|^{2}\label{eq:cor}
\]
\begin{equation}
\left\{4\left[\Psi(1)-\Psi(1/2)\right]-\mbox{ln}(x)-
\mbox{cc}-2\frac{\left(\partial_{\beta}+\partial_{\gamma}
\right)F(1/2,1/2;1,x)}{F(
1/2,1/2;1,x)}-\mbox{cc}\right\}
\end{equation}
where the logarithms and the $\Psi$ (Euler dilogarithm) appear due to
derivatives with respect to $c$, and $\beta$, $\gamma$ are the generic
arguments of the hypergeometric function $F(\alpha,\beta;\gamma,x)$. On the
other hand from the transformation formulas for $x\rightarrow 1-x$ one finds
\[
F(1/2,1/2;1,1-x)=-\frac{\Gamma(1)}{2\Gamma^{2}(1/2)}F(1/2,1/2;1,x)
\]
\begin{equation}
\left\{
4\left[\Psi(1)-\Psi(1/2)\right]-2\mbox{ln}(x)-4\frac{\left(\partial_{\beta}+
\partial_{\gamma}\right)F(1/2,1/2;1,x)}{F(1/2,1/2;1,x)}\right\}
\end{equation}
therefore one has that the part of (\ref{eq:cor}) in bracket is identical with
\begin{equation}
\left|F(1/2,1/2;1,x)\right|^{2}\left(\frac{F(1/2,1/2;1,1-x)}{F(
1/2,1/2;1,x)}+\mbox{cc}\right)
\end{equation}
This guarantees the identity of (\ref{eq:twistcor}) (for $k/N=1/2$)
 and (\ref{eq:cor}).

Besides the advantage that correlators of fields in the Kac table can be
computed without taking any limit, the main use of the fermionic formulation
is to provide expressions for the polymer operators with an {\bf odd} number of
legs. The most important is the one leg operator, for which one finds
\[
\left<\Phi_{1}(1)\Phi_{1}(2)\Phi_{1}(3)\Phi_{1}(4)\right>_{(12,34)+
(13,24)}
\]
\begin{equation}
\propto\left|
z_{12}z_{34}\right|^{3/8}\left|x(1-x)\right|^{3/8}
\left[F_{1/4}(x)\overline{F_{1/4}}(1-\overline{x})+F_{1/4}(1-x)
\overline{F}_{1/4}(\overline{x})\right]\label{eq:coroneleg}
\end{equation}
where the subscript $(12,34)+(13,24)$ means that one of the polymers connects
points $1$ to $2$ (resp. $1$ to $3$), the other one $3$ to $4$ (resp. $2$ to
$4$) (see figure 10). The above correlator is not invariant in $x\rightarrow
1/x$ since sources and sinks of polymers are distinguished. What one may call
the "full" one leg polymer operator is represented by
$\left|\sigma_{1/4}\right>$ $+$ $\left|\sigma_{3/4}\right>$ and its four point
function is obtained by adding (\ref{eq:coroneleg}) and the other function
obtained by exchanging $1$ and $2$.

Knowledge of the four point functions gives now access to the polymer operator
algebra.

Consider first the $L=4l+2$ legs operators that all belong to the Neveu
Schwartz sector.  Short distance product of these must give rise to operators
in the Neveu Schwartz sector again. This is confirmed by using the fusion rules
of the $\Phi_{1,3+2l}$ operators. One finds therefore
\begin{equation}
\Phi_{4l_{1}+2}.\Phi_{4l_{2}+2}\propto\sum_{l=0}^{l_{1}+l_{2}+1}\Phi_{4l+2}
\end{equation}
In a similar fashion consider the $L=4l$ operators in the Ramond sector. Short
distance product of these must give operators in the Neveu Schwartz sector.
this is confirmed by the fusion rules of $\Phi_{1,2+2l}$ operators
\begin{equation}
\Phi_{4l_{1}}.\Phi_{4l_{2}}\propto\sum_{l=0}^{l_{1}+l_{2}}\Phi_{4l+2}
\end{equation}
As far as operators with an odd number of legs are concerned we must use the
above analysis of the $Z_{4}$ sector and correlators. Calling $\Phi_{2l+1}$
 the full (in the above sense) $L=4l+1$ legs
operators one finds that product of two fields on the $Z_{4}$ sector gives rise
to fields both in the Neveu Schwarz and Ramond sector that is
\begin{equation}
\Phi_{2l_{1}+1}.\Phi_{2l_{2}+1}\propto\sum_{l=1}^{l_{1}+l_{2}+2}\Phi_{2l}
\end{equation}
The geometrical interpretation of these fusion rules is interesting and
discussed in figure 11.

\section{The Continuum Limit of Dilute Polymers}

Recall first that dilute polymers are obtained by considering a finite number
of self avoiding, mutually avoiding loops or open chains with a fugacity
$\mu^{-1}$ per unit length, where $\mu$ is the inverse effective connectivity
constant. We do not have  an "almost" soluble model to describe the physics of
dilute
polymers. The study of the $O(n),n\rightarrow 0$ phase transition
 \cite{SA87,PS80} suggests that there is a boson fermion "supersymmetry" in the
dilute phase such that $n=0$ is formally recovered by a cancellation between
the two species, and that this symmetry is broken in the dense phase. We know
on the other hand from the above analysis that the dense phase is described by
an $\eta,\xi$ system. This corresponds as well to taking only the fermionic
generators of the $gl(1,1)$ algebra. Restoring the boson fermion symmetry
should amount to using all the generators of this algebra. The most naive guess
would  then be to assume
 that the degrees of freedom of the continuum limit of dilute polymers
are $Gl(1,1)$ matrices, with a Wess Zumino action. As we commented in the
introduction this is not true. It is nevertheless instructive to carry out the
study of the $Gl(1,1)$ WZW model, as the related free field representation will
provide the right space to study dilute polymers, or twisted $N=2$
supersymmetry.
The Gaussian
 decomposition of a $Gl(1,1)$  element is
\begin{equation}
g=\left(\begin{array}{cc}
1&0\\
\xi&1
\end{array}\right)\left(\begin{array}{cc}
e^{\frac{i}{\sqrt{\kappa}}(\tilde{\phi}'-\frac{1}{2}\phi)}&0\\
0&e^{\frac{i}{\sqrt{\kappa}}(\tilde{\phi}'+\frac{1}{2}\phi)}
\end{array}\right)\left(\begin{array}{cc}
1&-\xi^{+}\\
0&1
\end{array}\right)
\end{equation}

 Working out the expression of the currents
$\partial^{\mu}gg^{-1}$ one  finds \cite{RS91}, after taking into account
 the various
quantum corrections, the action (where $\kappa$ is the level)
\begin{equation}
S=\frac{1}{\pi}\int d^{2}z\left[\partial\tilde{\phi}\ \overline{\partial}\phi-
\frac{i}{8\sqrt{\kappa}}{\cal R}\phi+\eta\overline{\partial}\xi+\ \mbox{
cc}\right]
\end{equation}
where the fields $\eta$ and $\tilde{\phi}$
 are obtained from $\partial\xi^{+}$ and
$\tilde{\phi}'$ by including quantum corrections, ${\cal R}$ is the scalar
curvature of the two dimensional metric. Corresponding to this
 action we have the
following propagator
\begin{equation}
<\phi(z)\tilde{\phi}(w)>=-\mbox{log}(z-w)
\end{equation}
while for the fermions (\ref{eq:fermionspropagator}) still holds. The stress
tensor reads
\begin{equation}
T(z)=-:\eta\partial\xi:-:\partial\phi\partial\tilde{\phi}:+\frac{i}
{2\sqrt{\kappa}}\partial^{2}\phi\label{eq:strten}
\end{equation}
In the follwing we set
\begin{equation}
\beta=\frac{1}{4\sqrt{\kappa}}
\end{equation}
The $Gl(1,1)$ WZW model \cite{RS91} turns out
to possess a number of features that are not compatible with our knowledge of
dilute polymers physics, like a spectrum not bounded from below\footnote{It is
interesting to notice that the $Gl(1,1)$ WZW model was also introduced to give
sense to a $n\rightarrow 0$ limit, but in the context of link invariants.}.
Nevertheless the above free field representation provides the correct fock
space to represent the dilute polymers. Indeed
 one recognizes in the stress tensor (\ref{eq:strten}) the stress tensor of
a twisted $N=2$ superconformal theory. Recall that
 the Coulomb gas representation of
ordinary $N=2$ superconformal theory as worked out for instance in
\cite{MSS89,I89,YZ87} involves the  degrees of freedom
$\psi,\tilde{\psi},\phi,\tilde{\phi}$, with the stress tensor
\begin{equation}
T_{N=2}(z)=-\frac{1}{2}:\left(\psi\partial\tilde{\psi}+\tilde{\psi}
\partial\psi
\right):-:\partial\phi\partial\tilde{\phi}:+
i\beta\left(\partial^{2}
\phi+\partial^{2}\tilde{\phi}\right)
\end{equation}
where now $\psi$ and $\tilde{\psi}$ are both fermions of weight $1/2$ and
propagators as in (\ref{eq:fermionspropagator}). The $U(1)$ current reads
\begin{equation}
J(z)=-:\psi\tilde{\psi}:+2i\beta\left(\partial\phi-\partial
\tilde{\phi}\right)
\end{equation}
with normalization
\begin{equation}
J(z)J(w)=\frac{c/3}{(z-w)^{2}}
\end{equation}
and central charge
\begin{equation}
c_{N=2}=3\left(1-\frac{1}{2\kappa}\right)
\end{equation}
The twisting is obtained \cite{EY90} by the substitution
\begin{equation}
T\rightarrow T+\frac{1}{2}\partial\ J
\end{equation}
leading to the result (\ref{eq:strten}) where we have relabelled $\psi$ by
$\eta$, $\tilde{\psi}$ by $\xi$ since these fields have now weights $1$ and
$0$. The central charge of the twisted theory is $c=0$ for any $\kappa$.
 We shall
now make the assumption that dilute polymers indeed are described by a twisted
$N=2$ theory, and investigate whether this is reasonable.  A few computations
of dimensions shows that the only possible level is
\begin{equation}
\kappa=3/4
\end{equation}
i.e. that the untwisted theory has $c=1,k=1$.

\subsection{Continuum Limit Partition Functions of Dilute Polymers}

\subsubsection{Spectral Flow and $\zeta$ Algebras in twisted and untwisted
$N=2$
   }

The characters of the  $N=2$ theory with central charge $c=1$ are
given for instance in \cite{D87,M87}. One finds the following results, where
the superscripts refer to the $U(1)$ charge and  dimension in the untwisted
theory
\begin{eqnarray}
\Xi_{NS}^{(0,0)}=\frac{1}{\eta}\sum_{n}q^{(6n)^{2}/24}\nonumber\\
\Xi_{NS}^{(\pm 1/3,1/6)}=\frac{1}{\eta}\sum_{n}q^{(6n+2)^{2}/24}
\end{eqnarray}
and similarly
\begin{eqnarray}
\Xi_{R}^{(\pm 1/6,1/24)}=\frac{1}{\eta}\sum_{n}q^{(6n-1)^{2}/24}\nonumber\\
\Xi_{R}^{(\pm 1/2,3/8)}=\frac{1}{\eta}\sum_{n}q^{(6n+3)^{2}/24}
\end{eqnarray}
We recall that, as everywhere in this work, the names Ramond and
Neveu Schwarz refer to boundary conditions {\bf in the plane}. When going to
the
cylinder, periodic and antiperiodic boundary conditions are exchanged in the
$N=2$ theory where the fermions have half integer conformal
 weights, but they remain
identical in the twisted theory where fermions acquire integer dimensions.

As emphasized in \cite{SS87} there is actually a continuous set of sectors for
the $N=2$ theory that are related to each other
by the spectral flow. For an arbitrary number $\zeta$ one can define the
$\zeta$ algebra for which the supersymmetry generators have boundary conditions
$G^{\pm}(z)=exp(\pm 2i\pi\zeta)G^{\pm}(e^{2i\pi}z)$, with
$\zeta=0$ for NS, $\zeta=\pm 1/2$ for R. They all have
 three irreducible representations
in our case. The corresponding transformation for weights and charges reads
\begin{equation}
(Q=x/6,h=x^{2}/4)\rightarrow (Q=(x-2\zeta)/6,h=(x-2\zeta)^{2}/24)
\end{equation}
To ensure modular invariance, $\zeta$ must be a rational number, $\zeta=k/N$.

Let us now consider the twisted theory. We first write the
 new relations
satisfied by the generators. We expand
\begin{eqnarray}
G^{+}(z)=\sum_{m}G^{+}_{m+1/2-k/N}z^{-m-2+k/N}\nonumber\\
G^{-}(z)=\sum_{m}G^{-}_{m+1/2+k/N}z^{-m-2-k/N}
\end{eqnarray}
and recall that $G^{+}$ acquires dimension $1$, $G^{-}$ dimension $2$. Then one
has
\begin{eqnarray}
\left[L_{m},L_{n}\right]&=&(m-n)L_{m+n}\nonumber\\
\left[L_{n},J_{m}\right]&=&-mJ_{m+n}-\frac{n(n+1)c}{6}\delta_{m+n}\nonumber\\
\left[J_{m},J_{n}\right]&=&c/3m\delta_{m+n}\nonumber\\
\left[L_{n},G^{+}_{m+1/2-k/N}\right]&=&-(m+1-k/N)G_{n+m+1/2-k/N}^{+}\nonumber\\
\left[L_{n},G^{-}_{m+1/2+k/N}\right]&=&(n-m-k/N)G_{n+m+1/2+k/N}^{-}\nonumber\\
\left\{G_{m+1/2-k/N}^{+},G_{n+1/2+k/N}^{-}\right\}&=&2\ L_{m+n+1}+2(n+1+k/N)
J_{m+n+1}\nonumber\\
&+&\frac{c}{3}[(m+1/2-k/N)^{2}-1/4]\delta_{m+n+1}\nonumber\\
\left[J_{n},G^{+}_{m+1/2-k/N}\right]&=&G^{+}_{n+m+1/2-k/N}\nonumber\\
\left[J_{n},G^{-}_{m+1/2+k/N}\right]&=&-G^{-}_{n+m+1/2+k/N}
\end{eqnarray}
Notice that, as in the dense polymers case, the $U(1)$ current $J$ is not a
(Virasoro) primary field anymore after twisting. The balance of charges is such
that
\begin{equation}
Q^{\dagger}=1/3-Q
\end{equation}
This balance can be decomposed in bosonic and fermionic balances (the fermionic
balance is the same as before)
\begin{equation}
Q_{F}^{\dagger}=1-Q_{F},Q_{B}^{\dagger}=-2/3-Q_{B}
\end{equation}
The twisting affects charge and dimension as follows
\begin{equation}
(Q= x/6,h=x^{2}/24)\rightarrow\ (Q= x/6,h=[(x- 1)^{2}-1]/24)
\end{equation}
A representation of the untwisted algebra is also a representation of the
twisted one, possibly reducible.We denote in the following the characters by
$\c
   hi_{l/N,k/N}^{(Q,h)}$
where as
in the dense case the labels $l/N,k/N$ stand for the fermions boundary
conditions, with $k/N=0$ for NS, $k/N=1/2$ for R. One finds
\begin{eqnarray}
\chi_{1/2,0}^{(0,0)}=\chi_{1/2,0}^{(1/3,0)}=\frac{1}{P}\sum_{n}
q^{[(6n+1)^{2}-1]/24}\nonumber\\
\chi_{1/2,0}^{(-1/3,1/3)}=\chi_{1/2,0}^{(2/3,1/3)}=\frac{1}{P}\sum_{n}
q^{[(6n+3)^{2}-1]/24}
\end{eqnarray}
where $P=\prod_{n=1}^{\infty}(1-q^{n})$ and
\begin{eqnarray}
\chi_{1/2,1/2}^{(-1/6,1/8)}=\chi_{1/2,1/2}^{(1/2,1/8)}=\frac{1}{P}\sum_{n}
q^{[(6n+2)^{2}-1]/24}\nonumber\\
\chi_{1/2,1/2}^{(1/6,-1/24)}=\frac{1}{P}\sum_{n}
q^{[(6n)^{2}-1]/24}\nonumber\\
\end{eqnarray}
We notice that the above sums are also those that apppear in the untwisted
theory, but NS and R sectors are exchanged. We notice also that for $k/N=0,1/2$
two of the three characters are in fact identical
. The characters $\chi_{0,1/2}$ are
obtained by definition by $\chi_{1/2,1/2}(\tau +1)$. For the doubly periodic
sector one finds $\chi_{0,0}^{(-1/3,1/3)}=\chi_{0,0}^{(2/3,1/3)}=0$ and
$\chi_{0,0}^{(0,0)}=\chi_{0,0}^{(1/3,0)}=1$.

By analogy with the dense polymer case we introduce also,
 in addition to the
NS and R sectors a $Z_{4}$  sector, with the characters in the twisted theory
\begin{eqnarray}
\chi^{(-1/12,5/96)}_{1/2,1/4}=\chi^{(5/12,5/96)}_{1/2,1/4}=\frac{1}{P}\sum_{n}
q^{[(6n+3/2)^{2}-1]/24}\nonumber\\
\chi^{(1/4,-1/32)}_{1/2,1/4}=\chi^{(1/12,-1/32)}_{1/2,1/4}=\frac{1}{P}\sum_{n}
q^{[(6n+1/2)^{2}-1]/24}\nonumber\\
\chi^{(-5/12,15/32)}_{1/2,1/4}=\chi^{(-5/12,15/32)}_{1/2,1/4}=\frac{1}{P}\sum_{n
   }
q^{[6n+5/2)^{2}-1]/24}
\end{eqnarray}
Notice that the three characters are now all different. Characters for other
values of $l/N$ are, by definition, obtained by successive modular
transformations $\tau\rightarrow\tau+1$. We also have
\begin{eqnarray}
\chi_{1/4,1/2}^{(-1/6,1/8)}=\frac{-i}{P}\sum_{n}(-i)^{n}
q^{[(6n-2)^{2}-1]/24}\nonumber\\
\chi_{1/4,1/2}^{(1/2,1/8)}=\frac{i}{P}\sum_{n}(-i)^{n}
q^{[(6n+2)^{2}-1]/24}\nonumber\\
\chi_{1/4,1/2}^{(1/6,-1/24)}=\frac{1}{P}\sum_{n}(-i)^{n}q^{[(6n)^{2}-1]/24}
\end{eqnarray}
and similar expressions for the $\chi_{1/4,0}$.

\subsubsection{The Sector with an Even Number of Polymers}

We now define, in a way similar to the dense case, the lattice partition
functions

\smallskip

{\bf Definition}

${\cal Z}_{AP(resp. PA, resp. AA)}$=Sum over configurations
 with an {\bf even}\footnote{Including zero} number of
non contractible dilute polymers, of total monomer length ${\cal L}$,  that
 cross $\omega_{2}$
(resp. $\omega_{1}$, resp. $(\omega_{1},\omega_{2})$
 an {\bf odd} number of times, with weight $2^{\mbox{number of
polymers}}$ $\mu^{-{\cal L}}$

\smallskip

where $\mu$ is the inverse connectivity constant. Contrarily to the dense case,
because dilute polymers do not occupy a finite fraction of the available space,
there is no free energy term as in (\ref{eq:freeenergy}), and one has to set
\begin{equation}
{\cal Z}_{PP}=1
\end{equation}
Let us define also the manifestly modular invariant quantity
\[
{\cal Z}^{e}=\left\{\mbox{Sum over configurations with an {\bf even}
number of non contractible}\right.
\]
\begin{equation}
\left.\mbox{ polymers, of total monomer length }{\cal L}
\mbox{, with weight }\ 2^{\mbox{number of polymers }}\ \mu^{-{\cal L}}\right\}
\end{equation}
One checks easily in that case that\footnote{Recall that the winding numbers of
a non contractible loop are coprimes}
\begin{equation}
{\cal Z}^{e}=\frac{1}{2}\left({\cal Z}_{AP}+{\cal Z}_{PA}+{\cal
Z}_{AA}-Z_{PP}\right)
\end{equation}
The continuum limit of these expressions was worked out in \cite{DS87}. The
same
expressions as the ones in subsection (3.1.2) apply, with the lattice Coulomb
gas coupling constant
$g=1/2$ replaced by $g=3/2$. The only difference is that the relation
(\ref{eq:relation}) is replaced by Euler identity
\begin{equation}
Z_{c}[3/2,2]-Z_{c}[3/2,1]=2
\end{equation}
and therefore
\begin{equation}
Z_{PP}=1
\end{equation}
Repeated use of the Jacobi identity provides the following results
\begin{eqnarray}
Z_{AP}=\frac{1}{2}\left(2\left|\chi_{1/2,0}^{(q,0)}\right|^{2}+
\left|\chi_{1/2,0}^{(q',1/3)}\right|^{2}
\right)\nonumber\\
Z_{PA}=\frac{1}{2}\left(2\left|\chi_{0,1/2}^{(q'',1/8)}\right|^{2}+\left|
\chi_{0,1/2}^{(1/6,-1/24)}\right|^{2}\right)\nonumber\\
Z_{AA}=\frac{1}{2}\left(2\left|\chi_{1/2,1/2}^{(q''',1/8)}\right|^{2}+\left|
\chi_{1/2,1/2}^{(1/6,-1/24)}\right|^{2}\right)\label{eq:Z2decomp}
\end{eqnarray}
Remarkably, this establishes that the boundary conditions of the continuum
fermions in the twisted $N=2$ theory are the same as those of lattice fermions
(although no discrete action is known to explicitely reproduce this) such that
a polymer loop is described either by a boson or a fermion, giving a total
weight zero for contractible ones, and zero or two for non contractible ones,
depending on the boundary conditions and the winding number. Adding the above
 results we find
\begin{equation}
Z_{AP}+Z_{PA}+Z_{AA}=Z_{c}[3/2,2]-1=Z_{c}[3/2,1]+1
\end{equation}
so that finally
\begin{equation}
Z^{e}=\frac{1}{2}Z_{c}[3/2,1]
\end{equation}
Remarkably also, the combination of partition functions for ${\cal Z^{e}}$
includes projection on odd fermion number.

That this special point of the Gaussian line was $N=2$ supersymmetric  was
observed by Friedan and Shenker \cite{FS88} and
S.K.Yang \cite{Y87}. Our point of view is of course a bit different
since we consider the above as a {\bf twisted} $N=2$ partition function.
{}From
this point of view, $Z^{e}$ is not a totally physical partition function
because it does not contain any ground states for the twisted theory, since we
have projection on odd fermion number. One can therefore decide to consider as
a physical partition function the one with projection on even fermion number
\begin{equation}
{\cal Z}^{phys}=\frac{1}{2}\left({\cal Z}_{AP}+{\cal Z}_{PA}+{\cal
Z}_{AA}+{\cal Z}_{PP}\right)
\end{equation}
with
\begin{equation}
{\cal Z}^{phys}\rightarrow \frac{1}{2}Z_{c}[3/2,2]
\end{equation}
In opposition to what happens in the case of say interacting around a face
models, adding a constant to the partition function does not change the
physics encoded in the fusion algebra and correlators. The combination of
sectors in the geometrical models seems to be merely a matter of convention.
Such an ambiguity did not occur in the dense case because $Z_{PP}$ vanished.
\footnote{Besides the physical interpretation of this result
 that was given above, one
can presumably explain it also  by considering that
the dense phase is in fact obtained by spontaneous breaking of supersymmetry of
a theory with vanishing Witten index. The index is two for dilute polymers}
Consider now,  as in the dense case

\[
{\cal Z}^{o}=\left\{\mbox{Sum over configurations with an {\bf} odd number of
noncontractible}\right.
\]
\begin{equation}
\left.\mbox{ polymers of total monomer length }{\cal L}
\mbox{, with weight }2^{\mbox{number of polymers  }}\mu^{-{\cal L}}\right\}
\end{equation}
whose continuum limit, evaluated by the Coulomb Gas mapping, is
\begin{equation}
Z^{o}=Z_{c}[3/2,1/2]-\frac{1}{2}Z_{c}[3/2,1]
\end{equation}
After some algebra one can also identify it with a twisted $N=2$ partition
function provided one introduces $Z_{4}$ sectors
\begin{eqnarray}
Z^{o}=\frac{1}{2}\left\{\sum_{l=0}^{3}\left[\left|\chi^{(q_{1},5/96)}
_{l/4,1/4}\right|^{2}+
\chi^{(q_{2},-1/32)}_{l/4,1/4}\left(\chi^{(q_{3},15/32)}_{l/4,1/4}\right)^{*}+
\mbox{cc}\right]+\right.\nonumber\\
\left.2\left|\chi^{(1/6,-1/24)}_{1/4,1/2}\right|^{2}+
\left|\chi^{(q',1/3)}_{1/4,0}
\right|^{2}+\chi^{(1/2,1/8)}_{1/4,1/2}\left(\chi^{(-1/6,1/8)}_{1/4,1/2}
\right)^{*}+\chi^{(1/3,0)}_{1/4,0}\left(\chi^{(0,0)}_{1/4,0}\right)^{*}+
\mbox{cc}\right\}\label{eq:Z4decomp}
\end{eqnarray}

The character decomposition of
\begin{equation}
{\cal Z}^{e+o}\rightarrow Z^{e+o}=Z_{c}[3/2,1/2]
\end{equation}
is finally obtained by adding the expressions
 (\ref{eq:Z2decomp}) and (\ref{eq:Z4decomp}).

It is also interesting to notice that the diagonal modular invariant, obtained
by summing the modulus square of each of the characters, turns out to be equal
to $Z_{c}[3/2,4]-1/2Z_{c}[3/2,2]$. Hence $Z_{c}[3/2,4]$ can also be expanded on
twisted $N=2$ characters.

\subsubsection{Operator Content of the Various Sectors}

We now discuss briefly the operator content of the various sectors.
Let us start with the antiperiodic (Ramond) sector. The ground state
$\left|\Sigma_{1/2}\right>$ has
conformal weight
\begin{equation}
h=-1/24
\end{equation}
To understand the physical meaning of this operator,
 imagine
a diluted version of the dense case. Hence consider again
configurations where polymers separate two points by encircling one of
them,with a factor of $2$. They are selected in the lattice
Coulomb gas by putting an
electric charge $-e_{0}$ ($e_{0}=1-g$) at both points such that polymer loops
encircling the punctures acquire weight $2$. The conformal dimension is
$h=-e_{0}^{2}/4g=-1/(8\times 3)=-1/24$. The content of the R sector is easily
found by decomposing each of the twisted $N=2$ characters on the $c=0$ Virasoro
characters using formulas in the appendix of \cite{DFSZ87}. Consider first the
representation $h=-1/24,Q=1/6$. One has
\begin{eqnarray}
\frac{q^{(12n)^{2}/24}}{\eta}=\sum_{k=n}^{\infty}\chi_{4k+2,3},
\ n\geq 0\nonumber\\
\frac{q^{(12n+6)^{2}/24}}{\eta}=\sum_{k=n}^{\infty}\chi_{4k+4,3},
\ n\geq 0
\end{eqnarray}
Therefore other Virasoro primary fields appear with weights
\begin{equation}
h_{4l}=h_{2l+2,3}=-\frac{1}{24}+\frac{3}{2}l^{2}
\end{equation}
They  all have even  degeneracy  (except for $l=0$) due to the $G^{\pm}$
symmetry. Their fermion number is equal to one (zero) for $l$ odd (even).
For instance the four polymers operator corresponds to
$G^{+}_{-2}\left|\Sigma_{1/2}\right>$ and
 $G^{-}_{-1}\left|\Sigma_{1/2}\right>$.
 The operators with $h=\overline{h}=h_{4l}$ ($l>0$) correspond to the
$L=4l$
legs operators. Their $U(1)$ charge is
$Q=1/6\pm L/4$.  For the representation $h=1/8,Q=-1/6$
 one uses
\begin{eqnarray}
\frac{q^{(12n-2)^{2}/24}}{\eta}=\sum_{k=n}^{\infty}\chi_{4k,1}+
\chi_{4k+2,2},\ n>0\nonumber\\
\frac{q^{(12n+4)^{2}/24}}{\eta}=\sum_{k=n}^{\infty}\chi_{4k+2,1}+
\chi_{4k+4,2},\ n\geq 0
\end{eqnarray}
These Virasoro primaries do not seem to correspond to any
 interesting geometrical
correlators. Actually they can be considered as subdominant operators for the
correlations of figure 4.

Consider now the NS sector. Notice that due to twisting we have now
\begin{eqnarray}
\left[L_{0},G^{\pm}_{\mp 1/2}\right]=0\nonumber\\
\left\{G^{+}_{-1/2},G^{-}_{1/2}\right\}=2L_{0}
\end{eqnarray}
Therefore every field {\bf but} the ground state with $h=0$ is twice
 degenerate.
 Let us discuss first the representation with $h=1/3,Q=-1/3$.
Using
\begin{eqnarray}
\frac{q^{(12n+3)^{2}/24}}{\eta}=\sum_{k=n}^{\infty}\chi_{4k+3,3}+
\chi_{4k+5,3}\nonumber\\
\frac{q^{(12n-3)^{2}/24}}{\eta}=\sum_{k=n}^{\infty}\chi_{4k+1,3}+\chi_{4k+3,3}
\end{eqnarray}
We find that the new Virasoro highest weight operators have weights
\begin{equation}
h_{4l+2}=h_{2l+3,3}=-\frac{1}{24}+\frac{3}{2}\left(l+\frac{1}{2}\right)^{2}
\end{equation}
They correspond to the $L=4l+2$ legs operators, with charge
$Q=1/6\pm L/4$. The representations $h=0,Q=0$ and
$h=0,Q=1/3$ both contain fields with integer dimensions only that do not seem
to have
any interesting meaning in terms of polymers.

Finally we consider the $Z_{4}$ sector. The Virasoro character expansion of
the twisted $N=2$ characters
 is straightforward since none of the dimensions belong to the Kac
table. For the representation $h=5/96,Q=-1/12$
one finds the set of conformal weights
\begin{eqnarray}
h_{4l+1}=-\frac{1}{24}+\frac{3}{2}\left(l+\frac{1}{4}\right)^{2}\nonumber\\
h_{4l+3}=-\frac{1}{24}+\frac{3}{2}\left(l+\frac{3}{4}\right)^{2}
\end{eqnarray}
which can formally be reproduced by the Kac formula
\begin{eqnarray}
h_{4l+1}=h_{2l+5/2,3}\nonumber\\
h_{4l+3}=h_{2l+7/2,3}
\end{eqnarray}
The ground state $\left|\Sigma_{1/4}\right>$ describes the one leg operator.
The three leg operator for instance can be obtained by acting with
$G^{+}_{-5/4}$ or $G^{-}_{-1/4}$.
The two other representations are combined in the partition
function. They
correspond to Virasoro primary fields with non vanishing spin, and some non
scalar polymer observables \cite{DS88}.

In summary the content in fuseau operators of the various sectors is once again
given by
\begin{eqnarray}
\mbox{Neveu Schwartz}:\ L=2\mbox{ mod }4\nonumber\\
\mbox{Ramond}:\ L=0\mbox{ mod }4\nonumber\\
Z_{4}:\ L=1,3\mbox{ mod }4
\end{eqnarray}
and the $U(1)$ charges are
\begin{equation}
\Phi_{L}\leftrightarrow Q=1/6\pm L/4
\end{equation}

\subsection{Correlators and Operator Algebra for Dilute Polymers}

The new Coulomb Gas made of the $\eta,\xi$ system and the
bosonic fields $\phi,\tilde{\phi}$ provides a much larger Fock space to
represent
the polymer theory than the usual Coulomb Gas a la Dotsenko Fateev, made of a
single bosonic field. This Fock space is quite big, and can be
restricted in many different ways (one of them corresponds for instance to the
$Gl(1,1)$ WZW model). Our task is to exhibit a subset of degrees of freedom for
which all four point functions can be built, and that is closed under operator
algebra.

Consider the vertex operators
\begin{equation}
V_{\alpha,\tilde{\alpha}}=\mbox{exp}
i\left(\tilde{\alpha}\phi+\alpha\tilde{\phi}\right)
\end{equation}
In the twisted theory they have conformal weight
\begin{equation}
h=\alpha(\tilde{\alpha}-2\beta)
\end{equation}
where we have set
\begin{equation}
\beta=\frac{1}{4\sqrt{\kappa}}=\frac{1}{\sqrt{12}}
\end{equation}
$h$ is invariant in the following three transformations
\begin{eqnarray}
\alpha\rightarrow 2\beta-\tilde{\alpha},\tilde{\alpha}\rightarrow
 2\beta-\alpha\nonumber\\
\alpha\rightarrow -\alpha,\tilde{\alpha}\rightarrow
4\beta-\tilde{\alpha}\nonumber\\
\tilde{\alpha}\rightarrow 2\beta+\alpha,\alpha\rightarrow\tilde{\alpha}-2\beta
\end{eqnarray}
The balance of (bosonic) charges to obtain a non vanishing correlator must be
\begin{equation}
\sum \alpha=0,\  \sum\tilde{\alpha}=4\beta
\end{equation}
Corresponding to a bosonic $U(1)$ charge
\begin{equation}
Q_{B}=2\beta(\alpha-\tilde{\alpha})=-2/3\label{eq:bosocharge}
\end{equation}
The three screening operators \cite{MSS89} of the untwisted theory can be used
as well in the twisted theory. Since they have vanishing total $U(1)$ charge
they are not affected by the twist. Introduce the charges
\begin{eqnarray}
\alpha_{+}=\tilde\alpha_{+}=2\beta\nonumber\\
\alpha_{-}=\tilde\alpha_{-}=-\frac{1}{2\beta}
\end{eqnarray}
The even screening operator reads
\begin{equation}
{\cal Q}_{e,1}=\oint dz
\left[i\partial(\tilde{\alpha}_{+}\phi-\alpha_{+}\tilde{\phi})+2\alpha_{+}
\tilde\alpha_{+}\xi\eta\right]e^{i(\alpha_{+}\tilde{\phi}+\tilde{\alpha}_{+}
\phi)}
\end{equation}
and the odd ones
\begin{eqnarray}
{\cal Q}_{o,1}=\oint dz \xi(z)e^{i\alpha_{-}\tilde{\phi}}\nonumber\\
{\cal Q}_{o,2}=\oint dz \eta(z)e^{i\tilde{\alpha}_{-}\phi}
\end{eqnarray}
where the coefficients in the prefactors are chosen to insure the vanishing of
the commutators of the ${\cal Q}$'s with the modes of $T,J,G^{\pm}$.

In the untwisted theory, the four point functions of the $N=2$ primary
fields can be built when the following quantization rule is obeyed
\begin{equation}
\alpha+\tilde{\alpha}=(1-n)\alpha_{+}-m\alpha_{-}
\end{equation}
They involve $(n-1)$ ${\cal Q}_{e,1}$ even screening operators, and $m$ of each
specie of odd screening operator ${\cal Q}_{o,1,2}$.

In the twisted theory we
shall compute the four point functions using the representation deduced from
the
untwisted case, that is
\begin{equation}
\left<V_{\alpha,\tilde{\alpha}}(1)V_{-\alpha,4\beta-\tilde{\alpha}}(2)
V_{\alpha,\tilde{\alpha}}(3)V_{\tilde{\alpha}-2\beta,\alpha+2\beta}(4)\right>
\end{equation}
The balance of charge in this correlator is
$\sum\alpha=\alpha+\tilde{\alpha}-2\beta$,
$\sum\tilde{\alpha}=\alpha+\tilde{\alpha}+6\beta$. Comparing with the
neutrality condition (\ref{eq:bosocharge}) we see
 that we have now to screen a net charge
$(\alpha+\tilde{\alpha}-2\beta,\alpha+\tilde{\alpha}+2\beta)$. Let us therefore
introduce the other even screening operator
\begin{equation}
{\cal Q}_{e,2}=\oint dz
\partial\phi\  e^{-2i\tilde{\alpha}_{+}\phi}
\end{equation}
Recall the expressions
\begin{eqnarray}
G^{+}=:-\sqrt{2}\xi\partial\phi+2\sqrt{2}i\beta\partial\xi :\nonumber\\
G^{-}=:-\sqrt{2}\eta\partial\tilde{\phi}+2\sqrt{2}i\beta\partial\eta:
\end{eqnarray}
and for the $U(1)$ current
\begin{equation}
J=:-\eta\xi-2i\beta\partial\tilde{\phi}+2i\beta\partial\phi
\end{equation}
One finds the following short distance expansions ($V$ is the vertex operator
in the integrand of ${\cal Q}_{e,2}$)
\begin{eqnarray}
G^{-}(z)V(w)&=&\sqrt{2}\partial_{w}\left(\frac{\eta
e^{-2i\tilde{\alpha}_{+}\phi(w)}}{z-w}\right)+\mbox{reg. terms}\nonumber\\
G^{+}(z)V(w)&=&0+\mbox{reg. terms}\nonumber\\
J(z)V(w)&=&2i\beta\partial_{w}\left(\frac{e^{-2i\tilde{\alpha}_{+}\phi(w)}}{z-w}
\right)\nonumber\\
T(z)V(w)&=&\partial_{w}\left(\frac{V(w)}{z-w}\right)
\end{eqnarray}
Ensuring commutation of ${\cal Q}_{e,2}$ with the twisted $N=2$ algebra. We
emphasize that ${\cal Q}_{e,2}$ is {\bf not}
 a screening operator for the untwisted
theory.

Operators in the Neveu Schwartz are represented by pure vertex operators. An
interesting example is the two polymer legs operator that can be represented
with
\begin{equation}
\Phi_{2}\leftrightarrow
(-\frac{\alpha_{+}+\alpha_{-}}{2},\frac{\tilde{\alpha}_{+}-
\tilde{\alpha}_{-}}{2})
\end{equation}
Its charge  is $Q=Q_{B}=-1/3$.
 One has $\alpha+\tilde{\alpha}$
$=-\alpha_{-}$. It therefore needs to be screened by  the pair ${\cal
Q}_{o,1},{\cal Q}_{o,2}$ plus ${\cal Q}_{e,2}$.

Operators in the Ramond sector are represented by the product of $\sigma_{1/2}$
and vertex operators. For the $h=-1/24$ operator that describes loops
separating two punctures, as explained above, one has the representation
\begin{equation}
\Phi_{0}\leftrightarrow
(-\alpha_{+}/2,\tilde{\alpha}_{+}/2)
\end{equation}
One has $\alpha+\tilde{\alpha}$ $=0$ such that screening by the operators
${\cal Q}_{e,1}$, ${\cal Q}_{e,2}$ is needed.

Finally operators in the $Z_{4}$ sector are represented by the product of a
vertex operator by a $\sigma_{1/4}$ twist. For the one leg operator one finds
\begin{equation}
\Phi_{1}\leftrightarrow
(\alpha_{+}/4-\alpha_{-}/2,-\tilde{\alpha}_{+}/4-\tilde{\alpha}_{-}/2)
\end{equation}
such that  $\alpha+\tilde{\alpha}$ $=-\alpha_{-}$ as for $\Phi_{2}$.
We expect once again that the position of the cuts for the $\eta,\xi$ fields
are in correspondence with the polymer lines.

Once these three basic four point functions are known, others can in principle
be deduced by acting with the twisted $N=2$ generators. For $L$ even they could
also be obtained by the Dotsenko Fateev representation using only one bosonic
field. For $L$ odd one needs the entire $N=2$ free field representation.

We can finally consider the polymer operator algebra. Consider first the
$L=4l+2$ operators that belong to the Neveu Schwartz sector. Upon fusion they
must give rise to  operators in NS again. This is confirmed by the fusion
rules of $\Phi_{2l+3,3}$ operators. One finds
\begin{equation}
\Phi_{4l_{1}+2}.\Phi_{4l_{2}+2}\propto\sum_{l=0}^{l_{1}+l_{2}+1}\Phi_{4l+2}+
\mbox{other operators with integer dimensions}
\end{equation}
For the $L=4l$ operators that belong to Ramond sector, short distance
expansuion
must produce fields in NS. This is confirmed by the fusion rules of
$\Phi_{2l+2,3}$. One finds
\begin{equation}
\Phi_{4l_{1}}.\Phi_{4l_{2}}\propto\sum_{l=0}^{l_{1}+l_{2}}\Phi_{4l+2}+
\mbox{other operators with integer dimensions}
\end{equation}
Finally  for the odd number of legs operators one has to use the $N=2$
representation and one finds that the $Z_{4}$ sector, upon fusion, branches to
both R and NS.
\begin{equation}
\Phi_{2l_{1}+1}.\Phi_{2l_{2}+1}\propto\sum_{l=1}^{l_{1}+l_{2}+2}\Phi_{2l}
+\mbox{other operators with integer dimensions}
\end{equation}

The closure of the full operator algebra for the polymer theory involving NS,
R,
and the $Z_{4}$ sector as coded in the partition functions $Z^{e},Z^{o}$ can be
checked as in \cite{MSS89}.

\section{Percolation and Related Problems}

\subsection{Percolation}

Recall that the percolation problem on say the square lattice is obtained by
choosing randomly edges to be occupied with a probability $p$ and empty with
probability $1-p$. The partition function is therefore one, but connected sets
of occupied bonds (clusters) possess non trivial geometrical properties.
Formally percolation is recovered by taking the $Q\rightarrow 1$ limit of the
$Q$ state Potts model. Contrary to the polymer problem, we do not have at hand
an almost soluble low temperature system to identify the degrees of freedom of
the theory. Let us therefore spend some time discussing what they could be. A
first possibility is to turn again to the medial graph, and make a polygon
decomposition of it with the same rules as in the spanning tree case. Instead
of a dense polymer, one gets a gas of polymer loops, each counted with a weight
one. One may think of reproducing this factor one by allowing $2{\cal N}+1$
bosons and $2{\cal N}$ fermions to propagate along each of these loops. In the
simplest case of ${\cal N}=1$ this provides a Lagrangian a la Parisi Sourlas
with $Sl(2,1)$ symmetry \cite{R91}. Since $sl(2,1)$ is isomorphic to $osp(2,2)$
   we are
entitled, as in the polymer case, to expect a continuum limit described by
twisted $N=2$ supersymmetry. However this introduction of fermions cannot be
entirely correct. In particular, choosing antoperiodic  boundary conditions for
the fermions, and formally taking ${\cal N}=1/2$ as in polymers,
 would lead to weights $3$ for non contractible loops, which cannot
be obtained in the Coulomb gas description with real charges. Another
alternative is to consider a field theory similar to the one of Isaacson and
Lubensky \cite{IL79} where the propagators draw directly the lattice clusters.
In that case changing the fermions boundary conditions would give a weight
$3$ to the non contractible clusters, ie a weight $\sqrt{3}$ to the non
contractible boundaries. This can be described in the lattice Coulomb Gas, and
this is the route we shall follow.

Consider therefore the percolation problem and its medial polygons on a
rectangle of the square lattice, and introduce as in the polymer case the
partition functions
\[
{\cal Z}^{1}_{++}=\left\{\mbox{Sum over dense coverings of the medial
 lattice with a gas
of loops}\right.
\]
\begin{equation}
\left.\mbox{each having weight one
and an even number of non contractible loops}\right\}
\end{equation}
and
\[
{\cal Z}^{1}_{-+(resp. +-,--)}=\left\{\mbox{Same as above,
 but non contractible loops that
cross}\right.
\]
\begin{equation}
\left.\omega_{1}
\left(\mbox{resp.}\omega_{2},\mbox{resp.}(\omega_{1},\omega_{2}\right)\mbox{
an odd
number of times having a weight } 3\right\}
\end{equation}
One finds, based on the lattice Coulomb Gas analysis,
\[
Z_{++}^{1}=\sum_{mm'\in Z}
\left(e^{2i\pi/3}\right)^{m\wedge m'}Z_{mm'}
\]
\[
Z_{-+}^{1}=\sum_{mm'\in Z}(-)^{m}
\left(e^{2i\pi/3}\right)^{m\wedge m'}Z_{mm'}
\]
\[
Z_{+-}^{1}=\sum_{mm'\in Z}(-)^{m'}
\left(e^{2i\pi/3}\right)^{m\wedge m'}Z_{mm'}
\]
\begin{equation}
Z_{--}^{1}=\sum_{mm'\in Z}(-)^{m+m'}
\left(e^{2i\pi/3}\right)^{m\wedge m'}Z_{mm'}
\end{equation}
After Poisson resummation one gets
\begin{equation}
Z^{1}=Z_{c}[2/3,6]-Z_{c}[2/3,2]
\end{equation}
Notice that Jacobi identity is conveniently written  in this case
\begin{equation}
Z_{c}[2/3,3]-Z_{c}[2/3,1]=2
\end{equation}
As was explained in \cite{DFSZ87}  the $Q$ state Potts model partition function
contains in addition a sector where only $Q-1$ colors are allowed on clusters
with "cross topology". It is therefore natural to consider also expressions
similar to the polymer case, where non contractible loops have weight zero or
two depending on the way they intersect the axis and the boundary conditions.
One finds
\begin{equation}
Z^{2}=Z_{c}[2/3,2]
\end{equation}
Summing both peaces gives rise to the result
\begin{equation}
Z=Z^{1}+Z^{2}=Z_{c}[2/3,6]=Z_{c}[3/2,4]
\end{equation}
This result is our proposal for the partition function of the percolation
problem in the context of twisted $N=2$ supersymmetry. As noticed in subsection
4.12, it
 is a diagonal twisted $N=2$ modular invariant involving the Neveu Schwartz,
Ramond and $Z_{4}$ sectors. The same analysis of the operator content as in the
dilute polymer case can be carried out. In the percolation problem, a first set
of interesting dimensions is provided by the analysis of boundaries of
clusters. The exponents for a $L$  hulls operator (see figure 12)
 are given by \cite{DS87bis}
\begin{equation}
h_{L}=\frac{4L^{2}-1}{24}
\end{equation}
For $L$ even they  belong to the Ramond sector. For $L$ odd they belong to
the Neveu Schwartz. Another set of interesting exponents are the thermal
operators (without obvious general geometrical meaning), whose dimensions are
\begin{equation}
h_{T_{n}}=\frac{(3n+1)^{2}-1}{24}
\end{equation}
They belong to Ramond for $n$ odd and Neveu Schwartz for $n$ even. Finally one
has the magnetic operators with dimensions
\begin{equation}
h_{H_{n}}=\frac{(6n-3)^{2}-4}{96}
\end{equation}
that belong to the $Z_{4}$ sector (see appendix A for more details on
percolation).

\subsection{Polymers at the Theta Point}

Recall that  polymers at the theta point are obtained by considering
 an attraction
 between monomers in addition to the self avoiding constraint. At a
certain temperature, these two competing interactions reach some equilibrium,
and the polymer is not as stretched as in the dilute case, not as compact as in
the dense case. As was argued in \cite{DS87bis} the statistics of such a
polymer
 is the same as the one of percolation perimeters. Introduce
therefore, by analogy with the dense and dilute cases, the partition functions
${\cal Z}^{e}$, ${\cal Z}^{o}$. One finds the continuum limits
\begin{equation}
Z^{e}=\frac{1}{2}Z_{c}[2/3,1]\ ,Z^{e+o}=Z_{c}[2/3,1/2]=Z_{c}[3/2,2]
\end{equation}
In that case only the Neveu Schwartz and Ramond sectors appear. The dimensions
of the $L$ legs operators are given by
\begin{equation}
h_{L}=\frac{L^{2}-1}{24}
\end{equation}
They belong to R for $L$ even, NS for $L$ odd. Notice that in this partition
function the ground state is twice degenerate, since the one leg
polymer operator has weight $h_{1}=0$. This is explained more naturally in the
other version of the problem provided by the Ising model. Consider indeed an
Ising model and its high temperature expansion. If we make analytic
continuation to the low temperature phase in a finite system, then take the
thermodynamic limit, we get again a model of dense
contours with weight one. The one
leg polymer operator becomes now the spin operator, and its vanishing dimension
occurs because of the finite spontaneous magnetization. We shall reconsider the
problem of the theta point and flow between various multicritical polymer
points in \cite{S91}.

\subsection{Brownian Walks}

Results on brownian motion are not very well established. Although the
exponents
$\nu=1/2$ (associated with a field $h=0$) and $\gamma=1$ (associated with
another $h=0$) are  easy to establish, the following exponents are non trivial.
Consider $M$ brownian walks connecting two points, that cannot intersect each
other. This defines a $M$ legs operator whose dimension is known from numerical
analysis \cite{DK88} (figure 13)
\begin{equation}
h_{M}=\frac{4M^{2}-1}{24}
\end{equation}
One notices that this exponent is the same as the one of the $L=2M$ legs
operator for tricritical
polymers, or as well $M$ percolation hulls.
 This suggests that the interesting geometrical objects are {\bf
boundaries}
of brownian walks, whose statistics would be the same as the one of
polymers at the theta point, or as well percolation hulls.
 The reason for this is unknown, but the result fits again in the
twisted $N=2$ framework. Moreover by choosing the partition function of the
doubly periodic sector to be the appropriate integer, the vanishing weights
occuring from the values of $\nu$ and $\gamma$ can also be included in the
theory.

\section{Conclusion}

In conclusion we would like first to comment on the "topological" nature of the
geometrical models we have considered \cite{W88}. Let us start with dense
polyme
   rs. In
that case we could use the fermionic operator ${\cal G}^{+}=\eta$ to define a
 BRS charge in
the periodic sector
\begin{equation}
Q_{BRS}=\oint \frac{dz}{2i\pi}\eta(z)
\end{equation}
It clearly satisfies $Q_{BRS}^{2}=0$.
 Moreover if we introduce  ${\cal G}^{-}$ $=$
$:\eta\xi\partial\xi:$, one has
\begin{equation}
\left\{Q_{BRS},{\cal G}^{-}\right\}=T
\end{equation}
such that the stress tensor is itself a BRS commutator. In that case, what is
usually considerd the "physical"
theory is trivial, ie $KerQ_{BRS}/ImQ_{BRS}$ is empty.
 The partition function for this
subset of states vanishes. Actually as is
readily seen from the $XX$ chain formulation, $Q_{BRS}$
 is the continuum limit of the
operator that creates polymers (that its square vanishes occurs because two
polymers cannot be created at the same place). Restricting to
$KerQ_{BRS}/ImQ_{BRS}$
means in lattice terms restricting to a sector without polymers, with vanishing
partition function $Z_{PP}=0$. Similarly if we consider dilute polymers in the
periodic sector and the charge
\begin{equation}
Q_{BRS}=\oint G^{+}
\end{equation}
one has, as is well known \cite{EY90}
\begin{equation}
\left\{Q_{BRS},G^{-}\right\}=T
\end{equation}
and the "physical" theory contains only the
identity. In that case again, it is not difficult to show how $Q_{BRS}$ is the
continuum limit of the operator that creates polymers. Restricting to $Ker
Q_{BRS}/ImQ_{BRS}$
 leaves only states without polymers, with partition function $Z_{PP}=1$
in agreement with $Ker Q_{BRS}/Im Q_{BRS}=Id$ in the conformal theory.

We therefore conclude that for statistical mechanics, physical states should
not be identified with the $Q_{BRS}$ cohomology.

The main points of this paper, besides providing a practical tool to study the
entire geometrical theories, are the connection between Parisi Sourlas
supersymmetry and twisted $N=2$, and the surprising unification of the standard
geometrical models  allowed by twisted $N=2$.
 All these models are described by $k=1$,
except the dense polymers where the symmetry is broken to an $\eta,\xi$ system.
 Dilute polymers and percolation
both involve  NS, R and a $Z_{4}$ sector. One corresponds to a diagonal, the
other to a non diagonal modular invariant. Tricritical polymers, contours in
the low temperature Ising model, and to a certain extent brownian motion,
involve only NS and R. Supersymmetry leads to
non trivial predictions as well. As an example we give in appendix A a
discussion of the backbone of percolation, together with the first conjecture
of
the exact value of its fractal dimension.
A natural extension of this work is the study of higher
values of $k$ in connection with multicritical geometrical problems \cite{S91}
. Another
would be to try to describe more complicated systems with Parisi Sourlas
Supersymmetry in the framework of twisted $N=2$.
Finally it is also mysterious how the Landau Ginzburg description of $N=2$
theories can be applied directly to geometrical problems \cite{S91}.

\bigskip

{\bf Acknowledgments}: I thank J.Descloizeaux who insisted on the incomplete
status of  polymer theory, and encouraged me to pursue the matter further. I
thank A.Leclair, G.Moore, N.Read and A.Zamolodchikov
for many very interesting conversations. I also thank P.Fendley
for discussions and collaboration at the preliminary stage of this work.

\pagebreak

{\bf Appendix A: Percolation Backbone}

\vspace{1cm}
The purpose of this appendix is to demonstrate that, besides its unifying
power, the recognition of $N=2$ supersymmetry in two dimensional  geometrical
models leads to new predictions as well. We consider here the percolation
problem. When an infinite cluster is formed at the percolation threshold, not
all its pieces can contribute to say electricity transport. One makes the
distinction between the useful part of the infinite cluster, the {\bf
backbone}, and the dangling ends. An important quantity is then the fractal
dimension of the backbone, which is a measure of the amount of useful edges,
and is smaller than the fractal dimension of the infinite cluster (the latter
equals $91/48$ in two dimensions). A nice theoretical \cite{H83}
way of studying the backbone is to consider operators whose correlation
function is defined as the sum over all clusters {\bf doubly} connecting them.
L
   et
us call in general, much like the $L$ legs operators for polymers, $\Psi_{L}$
the $L$ connectedness operator. One has obviously
\begin{equation}
D=2-2h_{L=2}
\end{equation}
So far no exact prediction for $h_{L}$
based on lattice Coulomb gas mapping or conformal
invariance had been made. Besides the difficulty of formulating the multiple
connectedness locally in lattice variables, the too vague status of the
conformal theory
description did not allow a search among the possible exponents. However  if we
now assume that all static observables for percolation
are described by a twisted $N=2$ theory, the choice becomes quite small
 (recall that modular invariance constraint requires the sectors to have
 $\eta$ a
rational number). Based on the order of magnitude of $D$ the only choice seems
to be
\begin{equation}
h_{2}=\frac{\eta(\eta+1)}{6},\ \eta=\frac{3}{4}
\end{equation}
ie
\begin{equation}
h_{2}=\frac{21}{96}
\end{equation}
This leads to
\begin{equation}
D=\frac{25}{16}
\end{equation}
in good agreement with numerical computations ($D\in 1.55-1.62$\cite{HHS84})
More generally our conjecture for the $L$ connectedness exponents is
\begin{equation}
h_{L}=\frac{\eta(\eta+1)}{6},\ \eta=\frac{2L-1}{4}
\end{equation}
It reproduces the correct results for $L=1,2$. It also corresponds to a field
 in the $Z_{4}$ sector, as expected since the multiple connectedness operators
all sit in the infinite cluster sector. Introducing the exponent
 $\beta=2\nu h$ with $\nu=4/3$  one finds
\begin{equation}
\beta_{L}=\frac{(2L-1)(2L+3)}{36}=L\beta_{1}+\nu \psi_{L}
\end{equation}
with
\begin{equation}
\psi_{L}=\frac{(L-1)(L+3/4)}{12}
\end{equation}
The result of $\epsilon$ expansion of \cite{H83} extended to two
dimensions is $\psi_{L}=16L(L-1)/49$. Like in the  polymers case,
 it has a form quite similar to  the conjectured exact one in two dimensions.

With a little optimism one may consider the numerical or experimental
verification of our $D=25/16$ prediction as a check of $N=2$ theory and
associated non
renormalization theorems.

\pagebreak

{\bf Appendix B: $S$ Matrices For Polymers}

\vspace{1cm}

The purpose of this appendix is to comment briefly on the use of the
supersymmetric formulation of polymers to study their off critical behaviour
(when perturbed by the thermal operator)
and the corresponding $S$ matrix. Indeed Zamolodchikov \cite{Z90} has written
an $S$ matrix for the $O(n)$ model involving $n$ particles, with the following
ansatz
\begin{equation}
S_{i_{1}i_{2}}^{j_{1}j_{2}}=S_{1}(\theta)\delta_{i_{1}}^{j_{2}}\delta_{i_{2}}
^{j_{1}}
+S_{2}(\theta)\delta_{i_{1}i_{2}}\delta^{j_{1}j_{2}}
\end{equation}
The factorization condition turned out to be equivalent to the follwing single
equation
\begin{eqnarray}
S_{1}(\theta)S_{1}(\theta+\theta')S_{2}(\theta')+S_{2}(\theta)S_{1}
(\theta+\theta')S_{1}(\theta')\nonumber\\
+S_{2}(\theta)S_{2}(\theta+\theta')S_{2}(\theta')
+nS_{2}(\theta)S_{1}(\theta+\theta')S_{2}(\theta')\nonumber\\
=S_{1}(\theta)
S_{2}(\theta+\theta')S_{1}(\theta')\label{eq:Scond}
\end{eqnarray}
The solution of (\ref{eq:Scond}) in the limit $n\rightarrow 0$ can be used to
compute some form factors. However in this limit
the $S$ matrix does not strictly make sense since in the point of view of
\cite{Z90} it has no degrees of freedom left onto which acting
. The above analysis suggests to give a sense
to the matrix elements extracted from (\ref{eq:Scond}) by introducing fermions.
Indeed let us consider a $S$ matrix acting in a graded space with N bosons, M
fermions, and the ansatz
\begin{equation}
S_{i_{1}i_{2}}^{j_{1}j_{2}}=S_{1}(\theta)(-)^{p(i_{1})p(i_{2})}
\delta_{i_{1}}^{j_{2}}\delta_{i_{2}}
^{j_{1}}+S_{2}(\theta)\delta_{i_{1}i_{2}}\delta^{j_{1}j_{2}}
\end{equation}
where $p$ is  the fermion number. One checks easily that, due to the
signs generated by application of $S_{13}$ when the state in the second space
is a boson, the graded factorization equation is satisfied if and only if the
equation (\ref{eq:Scond}) holds for
\begin{equation}
n=N-M
\end{equation}
Therefore the results of \cite{Z90} can actually be used to build a polymer $S$
matrix with as many bosons as fermions. It is likely that the present
formulation should allow application of the thermodynamic Bethe ansatz to
extract new interesting informations.  We also would like to comment that the
$S$ matrix in \cite{Z90} is algebraically equivalent to the sum of an identity
operator and a Temperley Lieb operator with parameter $\delta=n$. As far as the
computation of form factors is concerned, one can use as well the other
representation of the Temperley Lieb algebra provided by the 6 vertex model.
This means in particular that for the polymers,  another  alternative
to using bosons and fermions is to work with the Sine Gordon $S$ matrix. This
was also noticed in \cite{Sm91}.
This matrix exhibits precisely $gl(1,1)$ symmetry.

\pagebreak

{\bf Figure Captions}

\bigskip

Figure 1: The geometrical  operators $\Phi_{L}(z)$ are are defined by asking
tha
   t
$L$ legs of polymers emanate from $z$. The correlators
$\left<\Phi_{L}(z)\Phi_{L}(z')\right>$ are then obtained by summing over all
configurations where the $L$ legs describe self avoiding mutually avoiding
configurations starting form $z$ and enfding in $z'$.

Figure 2: A polygon decomposition of the medial lattice {\cal M} is obtained by
splitting each vertex in one of the two possible ways indicated.

Figure 3: A polygon decomposition of {\cal M} is dual to a subgraph of {\cal G}

Figure 4: The graphical representation of the Temperley Lieb algebra relations
is  precisely realized  in polymer systems.

Figure 5: Schematic representation of the subgraphs encountered for the
determinant of the Laplacian with one marked edge (indicated by a wiggly line).

Figure 6: An example of graph with a frustrated edge and the resulting
configurations made of one loop passing through the marked edge and dangling
ends attached to it.

Figure 7: The various subgraphs associated with the determinants given in
section 2.4

Figure 8: The geometrical interpretation of the Ramond sector ground state. The
frustration line connecting $z$ and $z'$ corresponds to antiperiodic boundary
conditions for fermions. Loops that intersect it an odd number of times (and
hence circle around one of the extremities) get a factor 2 instead of 0. Since
loops are dual to trees, they always occur in pairs.

Figure 9: The next geometrical excitation in the Ramond sector occurs when a
tree, ie a pair of loops, pinches the extremities $z$ and $z'$. This defines
the $4$ legs operator. In general the Ramond sector contains the $L=4l$ legs
operators.

Figure 10: Depending on the position of the cuts  for the
$Z_{4}$ twist in the $\eta,\xi$ system, different configurations with a pair of
polymers conecting four points can be obtained.

Figure 11: The fusion rules for operators with an even number of legs do not
correspond to naive addition of the number of legs. If $\Phi_{2l_{1}}$ and
$\Phi_{2l_{2}}$ are multiplied, one gets fields with a number of legs
$L=2l_{1}+2l_{2}+2$ modulo four, corresponding to additional "foldings" of some
of the legs.

Figure 12: Schematic representation of a 2 hulls operator for the  percolation
problem.

Figure 13: A 2 legs operator for brownian motion, with same exponent as a 2
hulls operator in percolation.

\end{document}